\documentclass[journal]{IEEEtran}

\usepackage{color,array,amsthm}
\usepackage{graphicx}

\usepackage{mathtools}
\usepackage[caption=false]{subfig}

\usepackage[utf8]{inputenc}
\usepackage{pgfplots}
\DeclareUnicodeCharacter{2212}{−}
\usepgfplotslibrary{groupplots,dateplot}
\usetikzlibrary{patterns,shapes.arrows}
\pgfplotsset{compat=newest}

\pgfplotsset{every axis/.append style={
	scaled x ticks = false,
	xticklabel style={
        /pgf/number format/.cd,
        fixed,
        1000 sep={},
        precision=0
        },
    legend style={font=\footnotesize},
	label style={font=\footnotesize},
    title style={font=\footnotesize},
	tick label style={font=\footnotesize},
	tick scale binop=\times
}}

\begin{document}

\title{Improving Galileo OSNMA Time To First Authenticated Fix} 

\author{Aleix~Galan, Ignacio~Fernandez-Hernandez, Wim~De~Wilde, Sofie~Pollin, Gonzalo~Seco-Granados%

\thanks{This research was partially funded by the Research Foundation Flanders (FWO) Frank de Winne PhD Fellowship, project number 1SH9424N (Aleix Galan).}%
\thanks{A.~Galan, I.~Fernandez-Hernandez and S.~Pollin are with KU Leuven. I.~Fernandez-Hernandez is also with the European Commission. Wim~De~Wilde is with Septentrio NV. G.~Seco-Granados is with UAB.}
\thanks{This work has been submitted to the IEEE for possible publication. Copyright may be transferred without notice, after which this version may no longer be accessible.}
}

\maketitle

\begin{abstract}
Galileo is the first global navigation satellite system to authenticate their civilian signals through the Open Service Galileo Message Authentication (OSNMA) protocol. However, OSNMA adds a delay in the time to obtain a first position and time fix, the so-called Time To First Authentication Fix (TTFAF). Reducing the TTFAF as much as possible is crucial to integrate the technology seamlessly into the current products. In the cases where the receiver already has cryptographic data available, the so-called \textit{hot start} mode and focus of this article, the currently available implementations achieve an average TTFAF of around 100 seconds in ideal environments. In this work, we explore the TTFAF optimizations available to general OSNMA capable receivers and to receivers with a tighter time synchronization than the required by the OSNMA Receiver Guidelines. We dissect the TTFAF process, describe the optimizations, and benchmark them in three distinct scenarios (open-sky, soft urban, and hard urban) with recorded real data. Moreover, we also evaluate the optimizations using the synthetic scenario from the official OSNMA test vectors. The first block of optimizations centers on extracting as much information as possible from broken sub-frames by processing them at page level and combining redundant data from multiple satellites. The second block of optimizations aims to reconstruct missed navigation data by the intelligent use of fields in the authentication tags belonging to the same sub-frame as the authentication key. Combining both optimization ideas improves the TTFAF substantially for all considered scenarios. We obtain an average TTFAF of 60.9 and 68.8 seconds for the test vectors and the open-sky scenario, respectively, with a lowest TTFAF of 44.0 seconds in both. Likewise, the urban scenarios see a drastic reduction of the average TTFAF between the non-optimized and optimized cases. These optimizations have been made available as part of the open-source OSNMAlib library on GitHub.
\end{abstract}

\begin{IEEEkeywords}
Global navigation satellite system, Galileo, OSNMA, Authentication, TTFAF optimization, OSNMAlib
\end{IEEEkeywords}

\section{INTRODUCTION}

G{\scshape lobal} Navigation Satellite Systems (GNSS) signals are vulnerable to interference, including the transmission of false GNSS-like signals, or spoofing. Adding cryptographic information to civil GNSS signals was proposed decades ago as a way to detect spoofing \cite{scott2003anti}, but it has taken time until its implementation. Meanwhile, several receiver-based anti-spoofing methods, such as signal power monitoring \cite{wesson2017gnss} or inertial systems \cite{tanil2017ins}, have been proposed. Finally, GNSS signals are gradually starting to provide cryptographic information.

Cryptographic techniques exploit the spoofer's ignorance of the cryptographic material when forging a signal. They can be applied to the spreading codes \cite{anderson2017chips} \cite{fernandez2023semi} or to the navigation data bits, which is known as Navigation Message Authentication (NMA). Although, in theory, a signal can still be replayed \cite{zhang2021protecting}, NMA facilitates the detection of such attacks \cite{humphreys2013detection} and provides very good protection against other common attack methods.

Galileo, the European GNSS, is the first GNSS to provide authentication to its civil signals and does so by implementing its own NMA-based protocol called OSNMA (Open Service Navigation Message Authentication). This protocol is the one used in this paper's research. It was proposed in the last decade \cite{fernandez2016navigation}, has been transmitted over the last years, and is expected to be launched operationally imminently \cite{gotzelmann2023galileo}. 

When adding OSNMA, receivers should not experience a degradation in accuracy or availability \cite{musumeci2023osnma}. However, the TTFF (Time To First Fix) will be impacted. This is mainly because OSNMA is based on TESLA (Timed Efficient Stream Loss-Tolerant Authentication) \cite{perrig2003tesla}, a delayed disclosure protocol, adapted to GNSS. The data and tags act as bit commitment, and the commitment is revealed later with the transmission of the symmetric TELSA key. A characteristic of delayed disclosure protocols is the requirement of an external loose time reference, and that they allow to use symmetric encryption algorithms. The symmetric encryption tags and keys are usually shorter than the signatures from a asymmetric encryption system, but their transmission increases the Time To First Authenticated Fix (TTFAF) with respect to TTFF \cite{fernandez2021analysis}. For Galileo, TTFF has been typically in the order of 30-60 seconds, although some recent improvements (the so-called \textit{I/NAV improvements}) in the navigation message will bring it to even lower values \cite{paonni2019improving}.

Specifically, we will focus on \textit{hot start} TTFAF, where the cryptographic information required to bootstrap the receiver is already known. Hereinafter, we will refer to TTFAF as hot start TTFAF. The OSNMA impact on TTFAF has been previously analyzed in the literature. Reference \cite{cucchi2022receiver} reaches an average TTFAF down to approximately 150 seconds including I/NAV improvements, and 170 seconds excluding them. In \cite{musumeci2023osnma}, the lowest case comparable to this work achieves 127 seconds. Reference \cite{gotzelmann2023galileo} achieves a lowest case of 120s, and \cite{hammarberg2024experimental} achieves 90s TTFAF. It is normal that these values vary, as they depend on the receiver implementation, which was not optimized to reduce TTFAF. We believe TTFAF optimization is relevant for potentially many OSNMA users, and is the focus of this paper. We propose several strategies to reduce OSNMA TTFAF down to 44 seconds in the lowest case, and test them in different environments.

To implement the proposed optimizations ,we used OSNMAlib \cite{osnmalib}, an open-source library that implements the OSNMA protocol which we developed in 2022 and maintained since then. As the library is written in Python, it is easy to modify and extend for research purposes, even though it might not be suitable for embedded purposes. 

OSNMAlib is not a receiver by itself, therefore it needs a GNSS receiver to track the satellites and decode the navigation data bits. For that purpose, we used Septentrio GNSS receivers (mosaic-X5 \cite{sept_mx5} and PolaRx5TR \cite{sept_polarx5tr}) to collect all the necessary data, which logging format is already integrated into our library.

The main contributions of this paper can be summarized as follows:
\begin{itemize}
    \item We propose two ideas to improve the TTFAF: page-level processing and COP-IOD optimization. The first approach is to extract partial information from broken sub-frames. The second idea goes even further and allows the reconstruction of missing navigation data by the innovative use of new OSNMA fields to improve TTFAF significantly.
    \item We validate these optimizations in three relevant scenarios using real data. The scenarios are diverse (open-sky, soft urban and hard urban) to show that the two proposed methods are very complementary and both ideas are needed to enable robust gains in all scenarios. We also evaluate the ideas using the official OSNMA test vectors.
    \item We analyze the OSNMA cross-authentication algorithm and the implications it has in the TTFAF when leveraging on the COP-IOD optimization.
    \item We provide an open-source implementation of the methods described in the paper in the OSNMAlib library.
\end{itemize}

The paper is organized as follows. The next section provides a general description of the OSNMA protocol and a brief summary of the OSNMAlib library. Then, the hot start TTFAF process and the proposed optimizations are detailed. This is followed by a description of the test scenarios used and, after, the test results are presented and discussed. The paper finalizes with the conclusions and further improvement ideas.


\section{GALILEO I/NAV, OSNMA AND OSNMAlib}

\subsection{Galileo I/NAV and OSNMA}

\begin{figure}
\centerline{\includegraphics[width=0.5\textwidth]{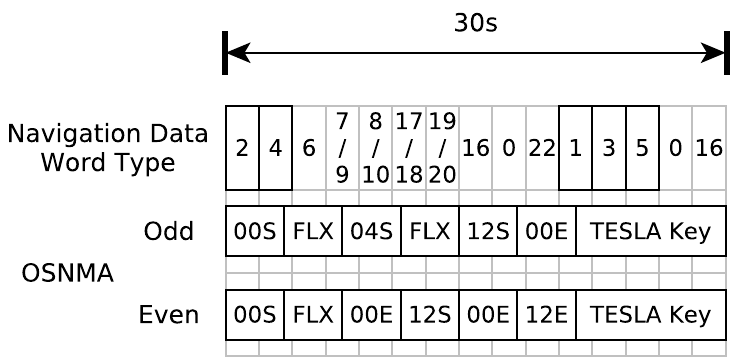}}
\caption{Operational configuration of Galileo navigation data and OSNMA data for one sub-frame. Some WTs alternate between even and odd sub-frames. The WTs with bold borders are used in the ADKD0 authentication. In this representation, the authentication tag size is 40 bits and the TESLA key size is 128 bits.}
\label{fig:ref_structure}
\end{figure}

Galileo OSNMA is transmitted in the I/NAV message, E1-B signal component \cite{Galileo_ICD}. The E1-B I/NAV message is composed of 30-second sub-frames of 15 two-second pages, each page including a Word Type (WT). WTs 1 to 5 contain the satellite ephemerides, ionosphere model, and health flags, and WTs 6 and 10 include time parameters (the latter shared with almanacs). There are other WTs, including only almanacs (WT 7 to 9) and spare words (WT 0). As part of the I/NAV improvements mentioned, Galileo has recently added new WTs: WT16 with a reduced ephemerides and WTs 17 to 20 with page recovery through Reed Solomon, which can be useful for OSNMA but we leave outside of our analysis for now. The WT order inside a sub-frame is represented in Fig. \ref{fig:ref_structure}.

OSNMA is inserted in Galileo’s E1-B page in a 40-bit field transmitted therefore every two seconds. As mentioned, OSNMA uses the TESLA protocol, with some variations and features such as key chain sharing across transmitting satellites and cross-authentication. The OSNMA 40-bit field is divided into the so-called HKROOT (Header and Root Key) section, of 8 bits, and the MACK (Message Authentication Codes and Key) section, of 32 bits. In this work we focus on the latter, which is the most relevant one for hot-start TTFAF.

In the MACK section, six truncated MACs, or \textit{tags}, are transmitted, preceding a key that authenticates the tags in the previous sub-frame (Fig. \ref{fig:ref_structure}). Each tag has 40 bits, and it incorporates a 16-bit \textit{tag-info} section, which encodes the satellite number the tag applies to and the type of authentication. At the moment, there are three types of tags, defined by the so-called ADKD (Authentication Data and Key Delay) parameter. ADKD0 and ADKD12 authenticate WTs 1 to 5, but ADKD12 with a key transmitted five minutes later to relax the receiver loose sync requirement, and ADKD4 authenticates the time (WT 6 and 10).

Due to system limitations, not all satellites can transmit OSNMA data at the same time. We refer to a satellite transmitting OSNMA as \textit{connected} and a satellite not transmitting OSNMA as \textit{disconnected}. To solve this limitation, OSNMA transmits cross-authentication tags that enable the authentication of disconnected satellites. The ADKD0 cross-authentication tag positions are named \texttt{00E} in Fig. \ref{fig:ref_structure}. There are also flex positions (\texttt{FLX}), which tag type is not predefined and needs to be verified at run-time, which are currently only used for ADKD0 cross-authentication tags. Therefore, there are 3 ADKD0 cross-authentication tags on each sub-frame.

For further details, a broad explanation of OSNMA is provided in \cite{fernandez2023galileo} and the full OSNMA specification can be found in the OSNMA SIS ICD \cite{OSNMA_ICD}. 

\subsection{OSNMA Time Synchronization}

For OSNMA to work securely, the receiver must know its synchronization accuracy with respect to Galileo System Time (GST). This requirement comes from the use of the TESLA protocol: when a TESLA chain key is disclosed, all the previous cryptographic material can be trivially forged. This implies that the receiver must have collected all the navigation data and associated tags before the appropriate TESLA chain key is revealed by the system.

The time synchronization requirement for OSNMA is defined as T\textsubscript{L} and set to 30 seconds: the time between the last bit of a tag and the first bit of the TESLA chain key authenticating it. A receiver that is not able to guarantee this T\textsubscript{L} cannot use ADKD 0 or 4 tags. However, it may use ADKD12 tags if the time synchronization is better than T\textsubscript{L} + 300 seconds.

A receiver has several strategies for obtaining the time synchronization with respect to the GST. If the receiver has no previous time information, it can retrieve the time from an external clock or from a secure Network Time Protocol (NTP) connection. In both cases, the time needs to be converted to estimate the GST and take appropriate measures to handle the associated uncertainty. If the receiver has already a time estimation and maintains it using an internal clock, the stability of the clock should be taken into account when verifying the time synchronization requirement. Further details and detailed procedures can be found in \cite{fernandez2020TL} and \cite{OSNMA_guidelines}.

Not complying with the synchronization requirement allows for arbitrary forgery attacks. For a normal OSNMA usage, T\textsubscript{L} is defined as 30 seconds. However, we will use tighter time synchronizations of 25 and 17 seconds for some of the optimizations described in this work.

\subsection{OSNMAlib}

OSNMAlib  \cite{osnmalib} \cite{osnmalib_conference} is an open-source library written in Python that implements the OSNMA protocol. The library can be integrated into existing receivers and applications to incorporate NMA into the PVT calculation. It can read the Galileo I/NAV pages from an input, store the navigation and authentication data, perform the verification operations, and report the status. The library supports cold start, warm start, and hot start procedures.

The input required for OSNMAlib to work is the navigation data bits from Galileo E1-B I/NAV message as nominal page, the Galileo System Time (GST) of the page transmission, and the Satellite Vehicle ID (SVID) to which the navigation data bits belong. Currently, OSNMAlib has the following input modules:

\begin{itemize}
    \item Septentrio SBF: Post-process files or live data in real-time from a Septentrio receiver in Septentrio Binary Format (SBF) if it contains the GALRawINAV block.
    \item u-blox UBX: Post-process files or live data in real-time from a u-blox receiver in UBX format if it contains the UBX-RXM-SFRBX message.
    \item GNSS-SDR: Process the output of the GNSS software-defined receiver project \cite{gnss-sdr-receiver} from a UDP socket.
    \item Galmon network: Connect to the Galmon network \cite{galmon} to process aggregated data from multiple receivers.
\end{itemize}

The library reports the OSNMA data received, the verification events, and the authenticated navigation data in chronological order. These logs also indicate when the receiver has enough authenticated data to calculate the first authenticated fix, together with the time elapsed since it started to process information. This logging option can be used to obtain the TTFAF value under different protocol configurations. Finally, the library also has a status logging every sub-frame in JSON format, which is useful for seeing the general state of OSNMA and extracting statistics about the scenario being processed. The status logging is used in the OSNMAlib webpage to display live information of the OSNMA protocol \cite{osnmalib_iclgnss}.


\section{PROPOSED TTFAF OPTIMIZATIONS}
For standard (unassisted) TTFF, the user needs to acquire and track signals, and decode the ephemerides (WTs 1 to 5) from at least 4 satellites, and time (WTs 6 and 10) from at least one. For TTFAF, the receiver also needs to receive the tags authenticating each of the above, and a TESLA key in the next sub-frame. Therefore, a delay is introduced.

For simplicity, we use the shorthand TTFAF to refer to TTFAF \textit{hot start}, i.e. when only the authentication tags and one TESLA key are needed, and the receiver has the cryptographic information to authenticate the key with a so-called root key already in its possession. The root key is expected to last for several months, hence it can be loaded to the receiver or reused from a previous execution. This is the standard operation mode and the focus of our paper.

Another start state is \textit{warm start}, where the receiver does not have the root key stored, but it has the public key needed to authenticate it. The receiver, then, needs to first retrieve the root key from the navigation data. The last start state is \textit{cold start}, where the receiver only has the Merkle Tree root hash needed to authenticate the public key in its possession. The public key is expected to last for several years and is transmitted every 6 hours.

The \textit{warm start} and \textit{cold start} are out of the scope of this paper since their TTFAF is bounded by other constraints, but the optimizations can still be applied retroactively once the receiver moves to the \textit{hot start} state.

\subsection{Page-Level Tag and Key Processing}\label{section:page_level_optimization}

At first glance, it may seem that OSNMA works at a per-satellite sub-frame level. The HKROOT is transmitted in numbered blocks that last one full sub-frame, and these sub-frame blocks need to be reordered to reconstruct the full message from multiple satellites. On the MACK side, a TESLA key is transmitted on every sub-frame to authenticate the tags of the previous sub-frame, and the tag order inside a sub-frame must be verified.

However, to optimize the OSNMA performance, a more granular approach should be taken. A Galileo sub-frame lasts 30 seconds and comprises 15 pages of 2 seconds each. Discarding all well-received pages of a sub-frame because the receiver missed one of them is not the most optimal method. The intelligent use of these pages can lead to the recovery of more OSNMA tags, as analyzed in \cite{damy_page_level_proc} with an older configuration of OSNMA that did not include flex tags. Therefore, to obtain a better TTFAF in hot start, we have implemented a page-level processing technique consisting of the following ideas.

The first idea is to extract tag sections from correctly received pages of partially corrupted sub-frames. For the secure use of OSNMA, the tags' order within a sub-frame must still be verified using the MAC look-up table or the MAC sequence value for the flex tags. Yet no flex tag may, in principle, be used in a sub-frame if the MAC sequence value or any other flex tag is missing. Consequently, a clear downside of having multiple flex tags in a MAC look-up table configuration is that this optimization will lose efficacy.

The second idea is to reconstruct the TESLA key by exploiting the diversity in the transmission. During a strong fading and poor visibility scenario, the receiver may not be able to fully retrieve the TESLA key from any satellite in view during one sub-frame. Nonetheless, that does not mean the TESLA key of that sub-frame is lost. Since all Galileo satellites transmit the same key during the same sub-frame, it may be possible to reconstruct the key using correctly received pages from different satellites.

\begin{figure}
\centerline{\includegraphics[width=0.5\textwidth]{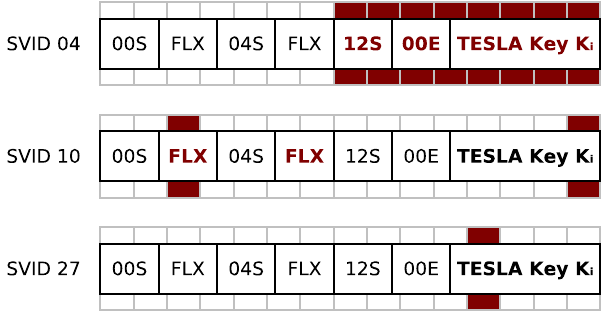}}
\caption{Missing a page (colored in red) affects part of the cryptographic data of the sub-frame, but the rest is still valid. Missing one flex tag means all flex tags are missed because their position cannot be verified. The TESLA key is the same for all satellites, so the receiver can reconstruct it by combining pages.}
\label{fig:page_level_processing}
\end{figure}

In Fig. \ref{fig:page_level_processing}, we show an example of how page-level processing helps extract valid cryptographic data. The tag sequence and key and tag sizes correspond to the OSNMA parameters transmitted during the OSNMA operational phase, illustrated in Fig. \ref{fig:ref_structure}. The figure depicts a sub-frame where satellite 04 moves out-of-sight, and the receiver misses the last few pages of the sub-frame. Nevertheless, the first four tags are perfectly useful. Satellite 10 misses a page corresponding to a flex tag, which affects the other flex tag, but the other four tags are valid. Both Satellite 10 and Satellite 27 miss a page of the key, so the sub-frame ends without any key fully received. However, the optimization is able to reconstruct the key because the satellites missed a different page.

Naturally, these optimizations are especially useful in scenarios with interference or fading where satellites are frequently out of sight. In a perfect open-sky scenario, only the low-elevation satellites entering or leaving the tracking horizon may have incomplete sub-frames.

\subsection{IOD Navigation Data Link} \label{section:iod_optimisation}

\begin{figure}[h]
\centerline{\includegraphics[width=0.5\textwidth]{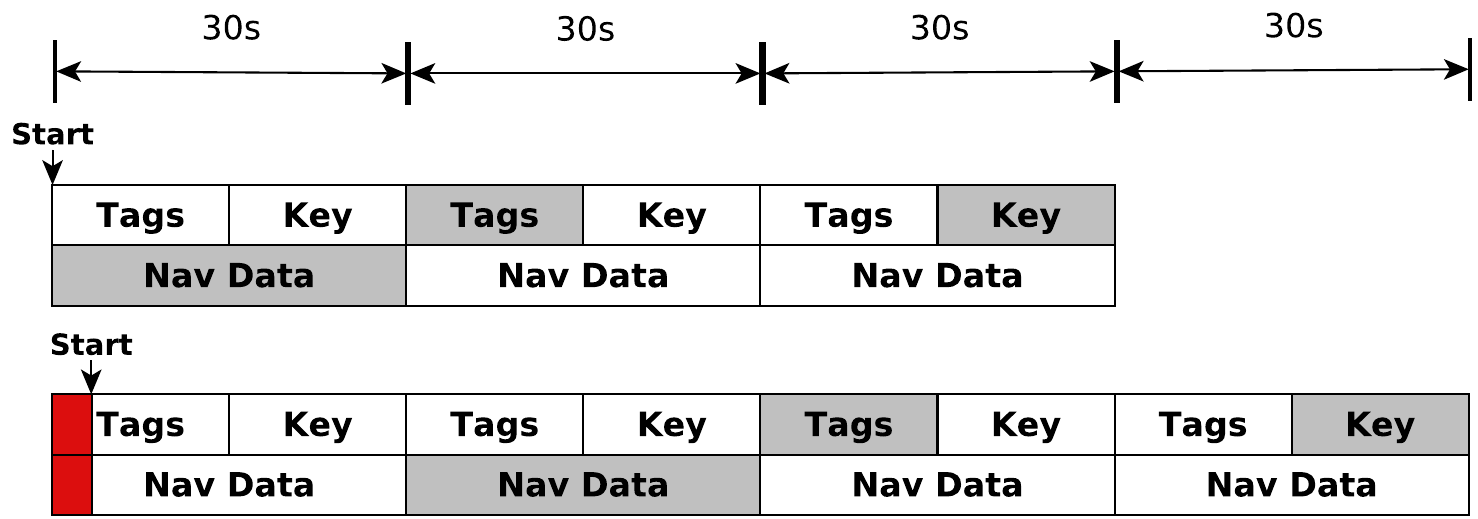}}
\caption{Depiction of the OSNMA navigation data authentication process without any optimization for one satellite. The top row indicates the OSNMA data received and the bottom the navigation data; the gray elements are used together to authenticate the navigation data. If the receiver does not start aligned with a sub-frame, it has to wait until the next.}
\label{fig:no_optimisation}
\end{figure}

Verifying the ADKD0 tags involves retrieving the navigation data, followed by the corresponding tag for this data in the subsequent sub-frame, and finally acquiring the TESLA key used for generating the tag in the third sub-frame. With this approach, a TTFAF of 90 seconds can be achieved as the lowest time, but if the receiver misses the first pages of the first sub-frame, it has to wait until the next one to start with the process, hence delaying the TTFAF to a maximum of 119 seconds. Fig. \ref{fig:no_optimisation} exemplifies these two cases for a single satellite. The top row indicates the OSNMA data, and the bottom row is the navigation data; both are transmitted in parallel. For any case between the lowest and the highest values, Fig. \ref{fig:theoretical_subframe} shows the TTFAF values depending on where the receiver starts in a sub-frame.

However, the ephemerides authenticated in ADKD0 change at a low rate and may be transmitted identically in several sub-frames. The data of multiple sub-frames can, therefore, be aggregated for authentication as long as it is the same. As discussed in the previous OSNMAlib paper \cite{osnmalib_conference} and the OSNMA receiver guidelines \cite{OSNMA_guidelines}, one way to reconstruct the navigation data from different sub-frames unambiguously is to use the Issue of Data (IOD) value transmitted in the I/NAV words, except WT 5, which does not have an IOD. Hence it must be assigned based on the IOD of other words of the sub-frame.

\begin{figure}[h]
\centerline{\includegraphics[width=0.5\textwidth]{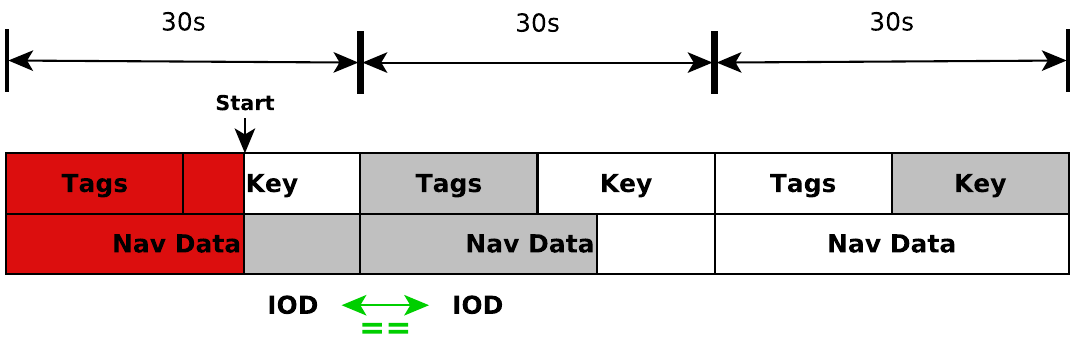}}
\caption{Depiction of the authentication data process for one satellite if the IOD of the data from the two sub-frames is the same. In this case, the missed navigation data can be retrieved from the next sub-frame.}
\label{fig:iod_optimisation}
\end{figure}

With this optimization, the lowest case occurs when the receiver starts processing navigation data immediately before WT 3 because it is the latest word containing the sub-frame IOD. WT 3 is transmitted 8 seconds before the end of the sub-frame, and we will have to wait for another sub-frame for the tags and another one for the key. Therefore, the lowest TTFAF is 60 seconds. If the navigation data does not change, the worst case occurs when the receiver starts immediately after WT 3 with a TTFAF of 97 seconds. A general example of this optimization is shown in Fig. \ref{fig:iod_optimisation}, and Fig. \ref{fig:theoretical_subframe} shows the TTFAF values for the IOD optimization as a function of sub-frame offset.

\subsection{Cut-Off Point Tag-Data Link}\label{section:cop_optimisation}

Originally, every tag included a 4-bit truncated IOD to link the tag with the data \cite{fernandez2016galileo}. However, the unpredictability of the IOD evolution in the system could lead to failed authentications if not appropriately handled. After some years in which the field was defined as 'Reserved', the last OSNMA specification has replaced this field by the 4-bit Cut-Off Point (COP) field \cite{OSNMA_ICD}. The COP indicates for how many sub-frames the navigation data authenticated with the tag has not changed. A value of 1 means that the authentication tag can only use navigation data from the previous sub-frame. A value of 15 (the maximum possible) indicates that the authentication tag can be verified using navigation data from the 15 previous sub-frames.

\begin{figure}
\centerline{\includegraphics[width=0.5\textwidth]{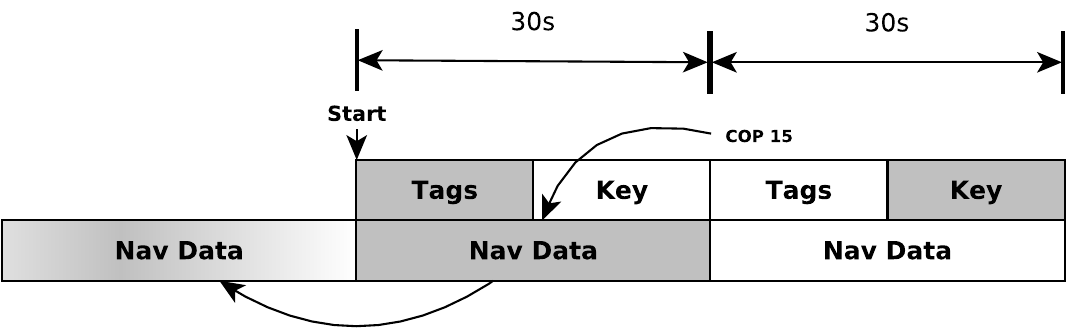}}
\caption{If there is a tag in the key sub-frame that authenticates the navigation data with a COP higher than 1, the data received in the tags sub-frame is the same as the previous sub-frame and can be used to authenticate the first tag.}
\label{fig:cop_optimisation}
\end{figure}

Although the original intention for the COP is to link the tag transmitting it with data from the previous sub-frames, we propose to use it to link other tags with the same data. With the traditional OSNMA approach, the receiver can never use the tags of the first sub-frame because the data transmitted in the previous sub-frame is unknown. However, this is the exact information given by the COP. If the navigation data has not changed, the COP of the tags in the key sub-frame will be greater than 1, indicating that the navigation data in the tags sub-frame is the same as in the prior sub-frame. Therefore, we can unambiguously link the tags received in the first sub-frame with the data of the first sub-frame (Fig. \ref{fig:cop_optimisation}).

Nevertheless, for this optimization to work, the receiver must get one tag in two consecutive sub-frames for the same navigation data. The tag received in the first sub-frame is used to authenticate the navigation data when the key is disclosed in the second sub-frame. The COP of the tag received in the second sub-frame is used to verify that the data received in the first sub-frame can be linked with the first sub-frame's tag.

\begin{figure}
\centerline{\includegraphics[width=0.5\textwidth]{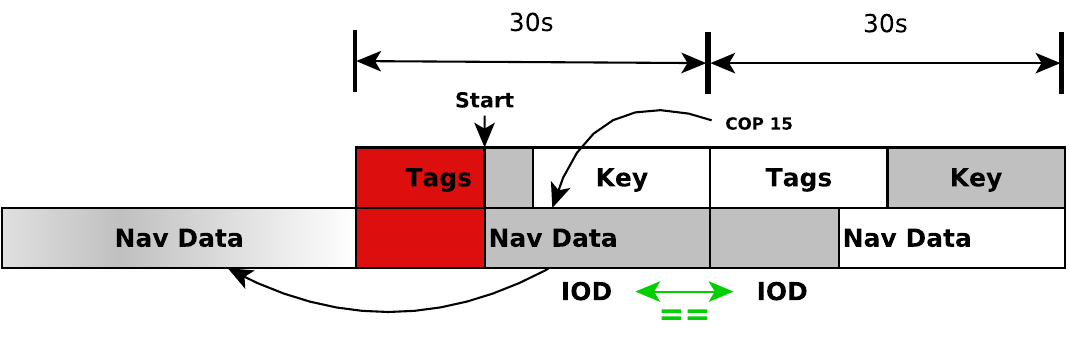}}
\caption{The IOD value can be used to bind navigation data of different sub-frames, and the COP value can be used to ensure the link between the navigation data and the authentication tag.}
\label{fig:cop_and_iod_optimisation}
\end{figure}

By using the COP value, it may seem that the previously discussed IOD optimization is no longer beneficial. However, both can be merged for even better TTFAF results. The same IOD logic to link navigation data from two sub-frames can be combined with the information provided by the COP as depicted in Fig. \ref{fig:cop_and_iod_optimisation}. The IOD links the navigation data from two sub-frames, and the COP shifts that data to the previous sub-frame, linking it with the tag.

The operations performed by a receiver implementing the COP and IOD optimization are the following:

\begin{enumerate}
    \item The receiver powers up in the middle of sub-frame $SF_j$, in time to get WTs 1, 3 and 5 from all Galileo satellites in view.
    \item At the end of $SF_j$, the receiver has also extracted a few cross-authentication tags from connected satellites. These tags authenticate navigation data transmitted at the previous sub-frame ($SF_{j-1}$), data that the receiver missed because it was not powered on.
    \item During the next sub-frame ($SF_{j+1}$), the receiver gets all the WTs from all satellites in view. For each satellite, if the IOD of these WTs is the same as the IOD of the WTs received at $SF_j$, the partial navigation data received at $SF_j$ can be fully reconstructed.
    \item Then, the receiver looks at the COP value of the authentication tags extracted during $SF_{j+1}$. If the COP value is greater than 1, it means that the reconstructed data for $SF_j$ is the same as the navigation data transmitted at $SF_{j-1}$ for the satellites targeted by the tags.
    \item At this moment, the receiver knows the navigation data transmitted at $SF_{j-1}$, has the tags to authenticate it (received at $SF_j$) and the TESLA key to verify them (received at $SF_{j+1}$). Therefore, it can proceed with the navigation data verification. 
\end{enumerate}
 
Combining the COP and the IOD, we obtain, in the lowest-case scenario, a TTFAF of 44 seconds on the even sub-frames or 46 seconds on the odd sub-frames. The position of the last cross-authentication tag in the tag sequence (Fig. \ref{fig:ref_structure}) defines the lowest possible TTFAF. If the navigation data does not change, the worst TTFAF is 73 seconds, when the receiver starts just after the last cross-authenticating tag.

We note that this optimization enables an acceptable forgery discussed in Section \ref{section:acceptable_forgeries}.

\section{FURTHER CONSIDERATIONS}

\subsection{Tighter Time Synchronization Requirement for the Optimizations}

The proposed optimizations in Sections \ref{section:iod_optimisation} and \ref{section:cop_optimisation} require a time synchronization with respect to GST lower than T\textsubscript{L} to work in a secure way. The OSNMA receiver must get all the authentication tags and navigation data before the corresponding TESLA chain key is disclosed to the system. With the optimizations, we are using navigation data words transmitted closer to the TESLA key than T\textsubscript{L}, thus requiring a tighter time synchronization. We define the time synchronization parameter T\textsubscript{S} as the maximum time synchronization the receiver can guarantee, independently of the method used to calculate it.

\begin{figure}[b!]
  \subfloat[No optimization. Navigation data used from the previous sub-frame.\label{fig:TL_base}]{\includegraphics[width=0.48\textwidth]{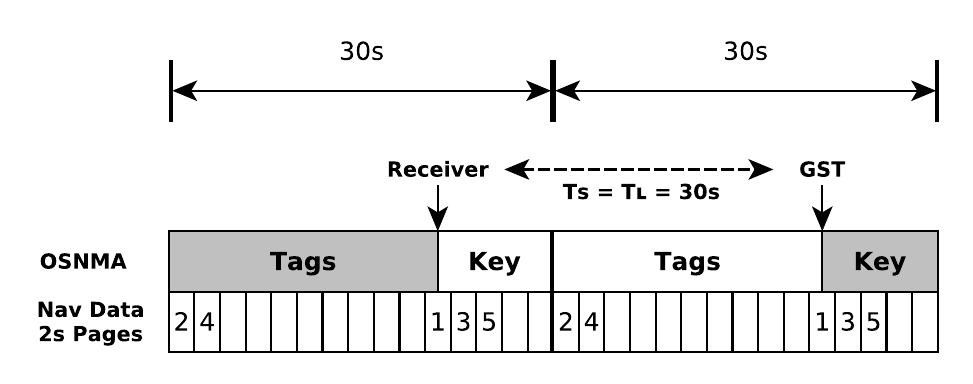}}\qquad
  \subfloat[IOD optimization.\label{fig:TL_iod}]{\includegraphics[width=0.48\textwidth]{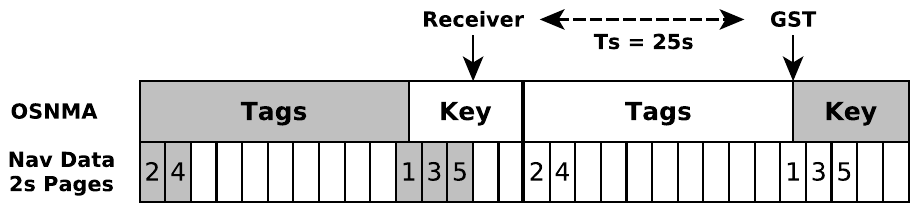}}\qquad
  \subfloat[COP and IOD optimization.\label{fig:TL_cop_iod}]{\includegraphics[width=0.48\textwidth]{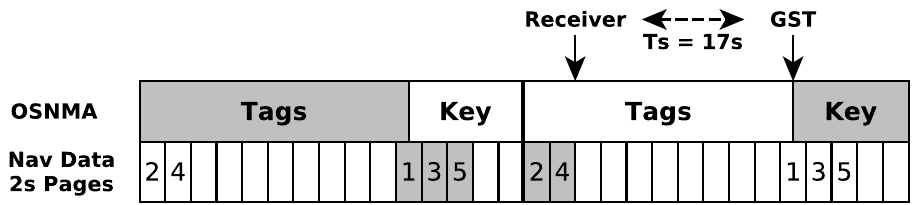}}

  \caption{The figure shows the calculation of the maximum time synchronization (T\textsubscript{S}) value for each of the optimizations. The downward arrows indicate the time as perceived by the receiver and the Galileo System Time (GST). The darker color indicates the elements used for the authentication.}
  \label{fig:TS_all}
\end{figure}

The different T\textsubscript{S} values are graphically shown in Fig. \ref{fig:TS_all} for one satellite. The time as perceived for the receiver and the GST time are indicated as downward arrows, and the material used for the authenticated is indicated in gray color.

For the IOD optimization described in Section \ref{section:iod_optimisation}, the receiver must be synchronized with the GST with a T\textsubscript{S} of 25 seconds to achieve maximum performance. The 25 seconds value corresponds to the time between the last bit of the last relevant navigation data word for ADKD0 (WT 5) transmitted in the tag sub-frame and the first bit of the TESLA key (Fig. \ref{fig:TL_iod}). The IOD optimization would work with a time synchronization of T\textsubscript{L}, but it could only link the WTs 2 and 4 with the navigation data of the previous sub-frame.

When using both the COP and the IOD to optimize the TTFAF as described in Section \ref{section:iod_optimisation}, the T\textsubscript{S} needed is 17 seconds. This value is the time between the last bit of the last relevant navigation data word for ADKD0 transmitted in the key sub-frame and the first bit of the TESLA key (Fig. \ref{fig:TL_cop_iod}). Although with a T\textsubscript{S} of one second it would also be possible to use the WT 1, we decided to discard the case because the WT is transmitted simultaneously as the key.

Aside from the not-optimized case, the rest of T\textsubscript{S} calculations depend on the key size, the tag size, and the number of tags transmitted on each sub-frame. For these results, we used the configuration transmitted during the data recording for this paper on December 03, 2023, which is the same configuration used for the operational phase of OSNMA (Fig. \ref{fig:ref_structure}).

Note that the navigation words order in the examples is extracted from the Galileo OS SIS ICD I/NAV Nominal Sub-Frame Structure for the E1-B signal \cite{Galileo_ICD}, which is only indicative. Also, a multi-frequency receiver capable of decoding the I/NAV stream from the E5b-I signal would get different values for the TTFAF and T\textsubscript{S}.

A receiver implementing these optimizations must take into account its T\textsubscript{S} and enable optimizations accordingly. In the case of OSNMAlib, the user may specify a time synchronization value different than T\textsubscript{L}, and the library will only use the tags and optimizations that are cryptographically secure for the value. Currently, the library does not support to change the T\textsubscript{S} after it starts to run, hence the receiver must ensure a lower value than the specified during the whole execution.

\subsection{Optimizations Theoretical Improvement}

The page-level tag and key processing optimization (Section \ref{section:page_level_optimization}) is scenario-specific, and its performance improvement will be determined by which pages the receiver misses. However, the TTFAF improvement of the tag-data link optimizations (Sections \ref{section:iod_optimisation} and \ref{section:cop_optimisation}) can be analyzed from a theoretical point of view.

For this exercise, we will analyze the theoretical TTFAF value depending on the start time of the receiver inside a sub-frame for 3 cases: the basic OSNMA without any optimization, the IOD optimization, which is already state of the art, and our newly proposed COP and IOD optimization. We will consider a single-frequency receiver (E1-B only) in an ideal open-sky scenario with 4 satellites in view, no pages lost, and no change in the navigation data.

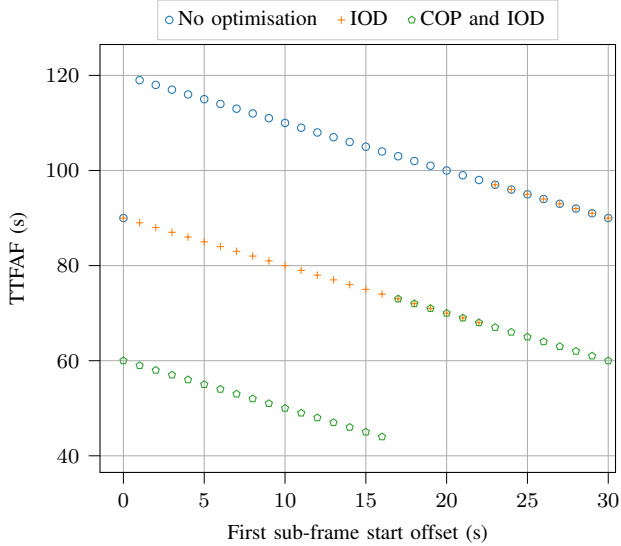
\begin{figure}
\centerline{
\begin{tikzpicture}

\definecolor{darkgray176}{RGB}{176,176,176}
\definecolor{darkorange25512714}{RGB}{255,127,14}
\definecolor{forestgreen4416044}{RGB}{44,160,44}
\definecolor{lightgray204}{RGB}{204,204,204}
\definecolor{steelblue31119180}{RGB}{31,119,180}

\begin{axis}[
legend cell align={left},
legend style={
    fill opacity=0.8,
    draw opacity=1,
    text opacity=1,
    draw=lightgray204,
    at={(0.5,1.01)},
    anchor=south,
    legend columns=3,
    /tikz/every even column/.append style={column sep=0.2cm}
},
tick align=outside,
tick pos=left,
x grid style={darkgray176},
xmajorgrids,
xmin=-1.45, xmax=30.45,
xtick style={color=black},
xlabel={First sub-frame start offset (s)},
y grid style={darkgray176},
ylabel={TTFAF (s)},
ymajorgrids,
ytick style={color=black},
]

\addplot [only marks, mark=o, mark size=1.4, steelblue31119180]
table {%
0 90
1 119
2 118
3 117
4 116
5 115
6 114
7 113
8 112
9 111
10 110
11 109
12 108
13 107
14 106
15 105
16 104
17 103
18 102
19 101
20 100
21 99
22 98
23 97
24 96
25 95
26 94
27 93
28 92
29 91
30 90
};
\addlegendentry{No optimisation}

\addplot [only marks, mark=+, mark size=1.4, darkorange25512714]
table {%
0 90
1 89
2 88
3 87
4 86
5 85
6 84
7 83
8 82
9 81
10 80
11 79
12 78
13 77
14 76
15 75
16 74
17 73
18 72
19 71
20 70
21 69
22 68
23 97
24 96
25 95
26 94
27 93
28 92
29 91
30 90
};
\addlegendentry{IOD}

\addplot [only marks, mark=pentagon, mark size=1.4, forestgreen4416044]
table {%
0 60
1 59
2 58
3 57
4 56
5 55
6 54
7 53
8 52
9 51
10 50
11 49
12 48
13 47
14 46
15 45
16 44
17 73
18 72
19 71
20 70
21 69
22 68
23 67
24 66
25 65
26 64
27 63
28 62
29 61
30 60
};
\addlegendentry{COP and IOD}
\end{axis}

\end{tikzpicture}}
\caption{Theoretical TTFAF values in an ideal scenario for the cases without optimization, with the IOD optimization, and with the COP-IOD optimization. The start time of the receiver within a sub-frame determines how long it will wait to get the first authenticated fix. The sub-frame start offset is relative to the first E1-B sub-frame from which the receiver starts decoding navigation data.}
\label{fig:theoretical_subframe}
\end{figure}

The results are shown in Fig. \ref{fig:theoretical_subframe} with the TTFAF value for the described optimizations as a function of offset of the first E1-B sub-frame for which the receiver starts getting navigation data.

\begin{figure}
\centerline{
\begin{tikzpicture}

\definecolor{darkgray176}{RGB}{176,176,176}
\definecolor{steelblue31119180}{RGB}{31,119,180}

\begin{axis}[
tick align=outside,
tick pos=left,
x grid style={darkgray176},
ylabel={Number of navigation data blocks},
xlabel={Seconds},
xmin=-1, xmax=25,
xtick style={color=black},
xtick={0,1,2,3,4,5,6,7,8,9,10,11,12,13,14,15,16,17,18,19,20,21,22,23,24},
xticklabel style={
    /tikz/.cd,
    rotate=80
    },
xticklabels={30,60,90,120,150,180,210,240,270,300,330,360,390,420,450,480,510,540,570,600,630,660,690,720,750+},
y grid style={darkgray176},
ymin=0, ymax=745.5,
ytick style={color=black}
]
\draw[draw=none,fill=steelblue31119180] (axis cs:-0.4,0) rectangle (axis cs:0.4,26);
\draw[draw=none,fill=steelblue31119180] (axis cs:0.6,0) rectangle (axis cs:1.4,4);
\draw[draw=none,fill=steelblue31119180] (axis cs:1.6,0) rectangle (axis cs:2.4,19);
\draw[draw=none,fill=steelblue31119180] (axis cs:2.6,0) rectangle (axis cs:3.4,1);
\draw[draw=none,fill=steelblue31119180] (axis cs:3.6,0) rectangle (axis cs:4.4,15);
\draw[draw=none,fill=steelblue31119180] (axis cs:4.6,0) rectangle (axis cs:5.4,4);
\draw[draw=none,fill=steelblue31119180] (axis cs:5.6,0) rectangle (axis cs:6.4,14);
\draw[draw=none,fill=steelblue31119180] (axis cs:6.6,0) rectangle (axis cs:7.4,4);
\draw[draw=none,fill=steelblue31119180] (axis cs:7.6,0) rectangle (axis cs:8.4,16);
\draw[draw=none,fill=steelblue31119180] (axis cs:8.6,0) rectangle (axis cs:9.4,3);
\draw[draw=none,fill=steelblue31119180] (axis cs:9.6,0) rectangle (axis cs:10.4,14);
\draw[draw=none,fill=steelblue31119180] (axis cs:10.6,0) rectangle (axis cs:11.4,4);
\draw[draw=none,fill=steelblue31119180] (axis cs:11.6,0) rectangle (axis cs:12.4,19);
\draw[draw=none,fill=steelblue31119180] (axis cs:12.6,0) rectangle (axis cs:13.4,4);
\draw[draw=none,fill=steelblue31119180] (axis cs:13.6,0) rectangle (axis cs:14.4,18);
\draw[draw=none,fill=steelblue31119180] (axis cs:14.6,0) rectangle (axis cs:15.4,4);
\draw[draw=none,fill=steelblue31119180] (axis cs:15.6,0) rectangle (axis cs:16.4,21);
\draw[draw=none,fill=steelblue31119180] (axis cs:16.6,0) rectangle (axis cs:17.4,4);
\draw[draw=none,fill=steelblue31119180] (axis cs:17.6,0) rectangle (axis cs:18.4,23);
\draw[draw=none,fill=steelblue31119180] (axis cs:18.6,0) rectangle (axis cs:19.4,710);
\draw[draw=none,fill=steelblue31119180] (axis cs:19.6,0) rectangle (axis cs:20.4,5);
\draw[draw=none,fill=steelblue31119180] (axis cs:20.6,0) rectangle (axis cs:21.4,1);
\draw[draw=none,fill=steelblue31119180] (axis cs:21.6,0) rectangle (axis cs:22.4,1);
\draw[draw=none,fill=steelblue31119180] (axis cs:22.6,0) rectangle (axis cs:23.4,0);
\draw[draw=none,fill=steelblue31119180] (axis cs:23.6,0) rectangle (axis cs:24.4,145);
\end{axis}

\end{tikzpicture}}
\caption{Transmission of the same navigation data from each satellite in view during a 24-hour recording. The majority of the time the same navigation data is transmitted for more than 600 seconds (20 sub-frames).}
\label{fig:data_change}
\end{figure}
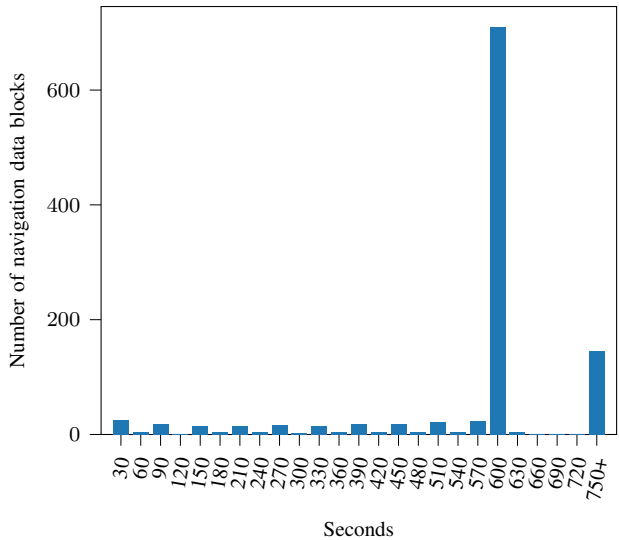

However, since the effectiveness of the optimizations is linked to the navigation data remaining the same between sub-frames, we have empirically analyzed how often the navigation data changes for each satellite. For this purpose we have used 24 hours of data from an open-sky receiver and calculated the duration of each block of navigation data (identified by the same IOD). The results shown in Fig. \ref{fig:data_change} clearly indicate that the majority of the time the navigation data gets updated after 600 seconds or more, which is 20 sub-frames.

For a more fine-graded approach, we have calculated the probability of the IOD optimizations working on any given sub-frame for a satellite and for a receiver. The distinction is that, while the navigation data may change for a satellite, the optimization will still work if it remains the same for at least 4 of them. Nonetheless, we have observed that the navigation data usually changes simultaneously for several satellites.

After analyzing the 24 hours of data recordings, the probability of the IOD optimizations working for a satellite on any given sub-frame is 96.28\%, and the probability of an OSNMA receiver being able to use the optimization on any given sub-frame is 97.78\% (Table \ref{tab:success_iod_optimization})

\begin{table}[h]
\renewcommand{\arraystretch}{1.3}
\caption{IOD optimizations success rate for a given sub-frame from a 24-hour recording. The optimizations do not work if the navigation data IOD changes. For the OSNMA receiver case, the optimization has to work for at least 4 satellites.}
\begin{center}
\begin{tabular}{|c|c|c|}
\hline
& \textbf{All Satellites} & \textbf{OSNMA Receiver} \\ \hline
Total Epochs                &  29239 & 2880  \\ \hline
Non-optimized Epochs        &  1087  & 64    \\ \hline
\textbf{Optimization Success (\%)}        &  \textbf{96.28} & \textbf{97.78} \\ \hline
\end{tabular}
\label{tab:success_iod_optimization}
\end{center}
\end{table}

\subsection{Acceptable Forgeries}\label{section:acceptable_forgeries}

There is a security consideration worth discussing with the COP link optimization described in Section \ref{section:cop_optimisation}: by using the COP value of the tags in the key sub-frame, we use unauthenticated information.

The receiver will accept 30 seconds old forged data in the event of a change in navigation data if the adversary modifies the COP value and transmits the previous data. In this case, the authentication will pass because the receiver reconstructs the navigation data block using correct data, but the applicability of the data will be 30 seconds off because it was not transmitted on that sub-frame.

However, this forging does not represent a risk in itself because, according to the Galileo System Definition Document, the navigation data has a validity of 4 hours without degrading the system performance \cite{Galileo_SDD}. Moreover, if there was no attack, the receiver would not be able to authenticate any data because the optimization does not work when the navigation data changes. The adversary is allowing the receiver to have an authenticated fix it would otherwise not have.

Nevertheless, the forging is detected later: when the tag containing the modified COP value is authenticated. If no pages are lost, this happens at the end of the next sub-frame (i.e. 30 seconds later). The adversary could try to jam the receiver and not allow it to get more navigation data, hence hiding the attack. However, the forged navigation data is correct, so the receiver could use it without added risks for as long as the data is valid.

Another method to avoid the 30-second misalignment on the data applicability time would be always to relate the data to the first sub-frame where a word is received and not the second. Therefore, in the case of accepting the forged data, the validity time would start in the first sub-frame (where the data was actually transmitted by the system) and not on the second sub-frame (where the data was transmitted by the adversary).

As a final note, to perform this forging attack the adversary must be able to replay the real Galileo signal and modify the navigation data fast enough to not fall behind the receiver's time synchronization. In such a scenario, general anti-replay techniques such as the use of partial correlations in the tracking loops \cite{gonzalo_partial_correlations} can also be applied to prevent the forging.


\section{SCENARIOS}

To evaluate the performance of the discussed optimizations, we recorded Galileo data in three relevant scenarios: a dynamic Hard Urban scenario, a dynamic Soft Urban scenario, and a static Open-Sky scenario. Additionally, we have also processed configuration 2 of the official OSNMA test vectors \cite{OSNMA_guidelines} because it contains the same tag sequence as the live transmitted data. For the dynamic recordings, we used a Septentrio mosaic-X5 with firmware version 4.14.0. For the static Open-Sky scenario, we used a Septentrio PolaRx5TR with firmware version 5.5.0.

The data recordings are saved in Septentrio Binary Format (SBF). This format contains the \textit{GalRawINAV} block with all the information needed to post-process the files with OSNMAlib (Galileo I/NAV message bits, SVID, and receiver GST). The recordings, containing all the GNSS logged information and format definition, are available in \cite{gnss_data}.

\subsection{Hard Urban Scenario: Brussels, European District}

\begin{figure}
  
  \subfloat[Trajectory followed.\label{fig:map_eu_district}]{\centerline{\includegraphics[width=0.5\textwidth,trim={8cm 10cm 5cm 10cm},clip]{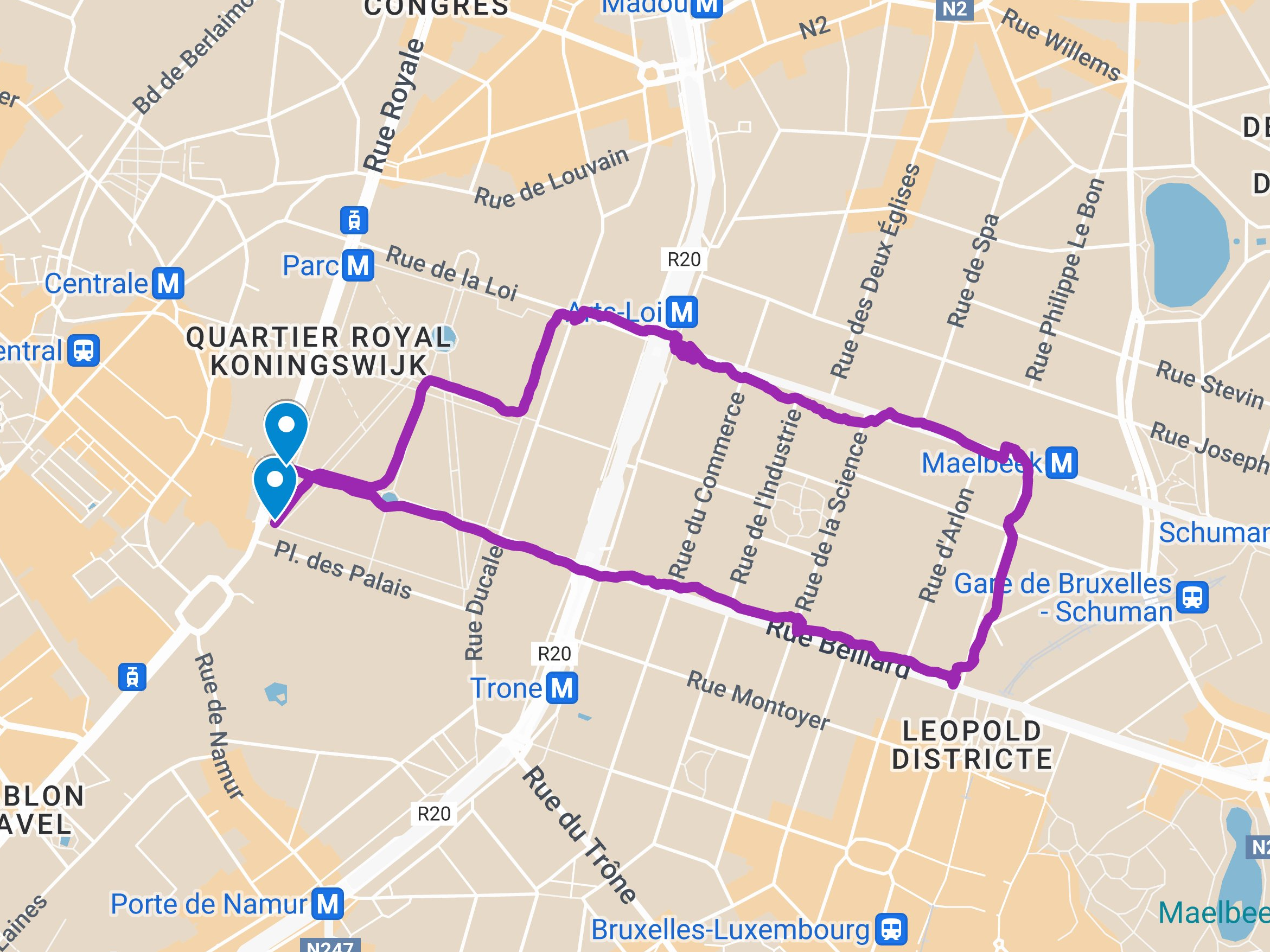}}}\qquad 
  \vspace{0.2cm}
  \subfloat[Galileo satellites tracked and OSNMA connected.\label{fig:satellites_eu_district}]{\centerline{
\begin{tikzpicture}

\definecolor{darkgray176}{RGB}{176,176,176}
\definecolor{lightgray204}{RGB}{204,204,204}

\begin{axis}[
legend cell align={left},
legend style={
    fill opacity=0.8,
    draw opacity=1,
    text opacity=1,
    draw=lightgray204,
    at={(0.5,1.01)},
    anchor=south,
    nodes={scale=0.8},
    legend columns=2,
    /tikz/every even column/.append style={column sep=0.2cm}
    },
tick align=outside,
tick pos=left,
x grid style={darkgray176},
xmajorgrids,
xmin=35304, xmax=37416,
xtick style={color=black},
y grid style={darkgray176},
ylabel={SVID},
ymajorgrids,
ymin=-0.35, ymax=7.35,
ytick style={color=black},
ytick={0,1,2,3,4,5,6,7},
yticklabels={4,5,9,15,24,31,34,36},
xlabel={Time of Week (s)},
width=0.47\textwidth,
height=4.5cm,
]

\addlegendimage{only marks, orange, mark=*, mark size=0.7, mark options={solid}}
\addlegendentry{Satellite tracked}
\addlegendimage{only marks, black!30!green, semithick, mark=triangle, mark size=1.25, mark options={rotate=180}}
\addlegendentry{Satellite OSNMA connected}

\addplot [orange, mark=*, mark size=0.7, mark options={solid}, only marks, forget plot]
table {%
35460 0
35490 0
35520 0
35580 0
35610 0
35670 0
35730 0
35760 0
35790 0
35880 0
35910 0
35940 0
35970 0
36000 0
36030 0
36060 0
36090 0
36120 0
36180 0
36330 0
36360 0
36390 0
36420 0
36450 0
36480 0
36540 0
36570 0
36630 0
36660 0
36750 0
36780 0
36810 0
36840 0
36870 0
36900 0
37020 0
37050 0
37110 0
37140 0
37170 0
37200 0
37230 0
37290 0
37320 0
};
\addplot [semithick, black!30!green, mark=triangle, mark size=1.25, mark options={rotate=180}, only marks, forget plot]
table {%
35460 0
35490 0
35520 0
35580 0
35610 0
35670 0
35730 0
35760 0
35790 0
35880 0
35910 0
35940 0
35970 0
36000 0
36030 0
36060 0
36090 0
36120 0
36180 0
36330 0
36360 0
36390 0
36420 0
36450 0
36480 0
36540 0
36570 0
36630 0
36660 0
36750 0
36780 0
36810 0
36840 0
36870 0
36900 0
37020 0
37050 0
37110 0
37140 0
37170 0
37200 0
37230 0
37290 0
37320 0
};
\addplot [orange, mark=*, mark size=0.7, mark options={solid}, only marks, forget plot]
table {%
35400 1
35430 1
35460 1
35490 1
35520 1
35550 1
35580 1
35610 1
35640 1
35670 1
35700 1
35730 1
35760 1
35790 1
35820 1
35850 1
35880 1
35910 1
35940 1
35970 1
36000 1
36030 1
36060 1
36090 1
36120 1
36150 1
36180 1
36210 1
36240 1
36270 1
36300 1
36330 1
36360 1
36390 1
36420 1
36450 1
36480 1
36510 1
36540 1
36570 1
36600 1
36630 1
36660 1
36690 1
36720 1
36750 1
36780 1
36810 1
36840 1
36870 1
36900 1
36930 1
36960 1
36990 1
37020 1
37050 1
37080 1
37110 1
37140 1
37170 1
37200 1
37230 1
37260 1
37290 1
37320 1
};
\addplot [orange, mark=*, mark size=0.7, mark options={solid}, only marks, forget plot]
table {%
35400 2
35430 2
35460 2
35490 2
35520 2
35550 2
35580 2
35610 2
35640 2
35670 2
35700 2
35730 2
35760 2
35790 2
35820 2
35850 2
35880 2
35910 2
35940 2
35970 2
36000 2
36030 2
36060 2
36090 2
36120 2
36150 2
36180 2
36210 2
36240 2
36270 2
36300 2
36330 2
36360 2
36390 2
36420 2
36450 2
36480 2
36510 2
36540 2
36570 2
36660 2
36690 2
36720 2
36750 2
36780 2
36810 2
36840 2
36870 2
36900 2
36930 2
36960 2
36990 2
37020 2
37050 2
37080 2
37110 2
37140 2
37170 2
37200 2
37230 2
37260 2
37290 2
37320 2
};
\addplot [semithick, black!30!green, mark=triangle, mark size=1.25, mark options={rotate=180}, only marks, forget plot]
table {%
35400 2
35430 2
35460 2
35490 2
35520 2
35550 2
35580 2
35610 2
35640 2
35670 2
35700 2
35730 2
35760 2
35790 2
35820 2
35850 2
35880 2
35910 2
35940 2
35970 2
36000 2
36030 2
36060 2
36090 2
36120 2
36150 2
36180 2
36210 2
36240 2
36270 2
36300 2
36330 2
36360 2
36390 2
36420 2
36450 2
36480 2
36510 2
36540 2
36570 2
36660 2
36690 2
36720 2
36750 2
36780 2
36810 2
36840 2
36870 2
36900 2
36930 2
36960 2
36990 2
37020 2
37050 2
37080 2
37110 2
37140 2
37170 2
37200 2
37230 2
37260 2
37290 2
37320 2
};
\addplot [orange, mark=*, mark size=0.7, mark options={solid}, only marks, forget plot]
table {%
35430 3
35460 3
35490 3
35520 3
35550 3
35580 3
35610 3
35640 3
35670 3
35700 3
35730 3
35790 3
35820 3
35910 3
35940 3
36000 3
36030 3
36060 3
36090 3
36120 3
36390 3
36450 3
36540 3
36570 3
36600 3
36690 3
36750 3
36780 3
36900 3
36930 3
36960 3
36990 3
37080 3
37110 3
37140 3
37170 3
37200 3
37230 3
37260 3
37290 3
37320 3
};
\addplot [semithick, black!30!green, mark=triangle, mark size=1.25, mark options={rotate=180}, only marks, forget plot]
table {%
35430 3
35460 3
35490 3
35520 3
35550 3
35580 3
35610 3
35640 3
35670 3
35700 3
35730 3
35790 3
35820 3
35910 3
35940 3
36000 3
36030 3
36060 3
36090 3
36120 3
36390 3
36450 3
36540 3
36570 3
36600 3
36690 3
36750 3
36780 3
36900 3
36930 3
36960 3
36990 3
37080 3
37110 3
37140 3
37170 3
37200 3
37230 3
37260 3
37290 3
37320 3
};
\addplot [orange, mark=*, mark size=0.7, mark options={solid}, only marks, forget plot]
table {%
35400 4
35520 4
35550 4
35580 4
35880 4
35940 4
36060 4
36270 4
36300 4
36330 4
36420 4
36450 4
36510 4
36540 4
36630 4
36660 4
36690 4
36720 4
36750 4
36780 4
36810 4
36900 4
36930 4
36960 4
36990 4
37020 4
37050 4
37080 4
37110 4
37200 4
37230 4
37260 4
37290 4
37320 4
};
\addplot [semithick, black!30!green, mark=triangle, mark size=1.25, mark options={rotate=180}, only marks, forget plot]
table {%
35400 4
35520 4
35550 4
35580 4
35880 4
35940 4
36060 4
36270 4
36300 4
36330 4
36420 4
36450 4
36510 4
36540 4
36630 4
36660 4
36690 4
36720 4
36750 4
36780 4
36810 4
36900 4
36930 4
36960 4
36990 4
37020 4
37050 4
37080 4
37110 4
37200 4
37230 4
37260 4
37290 4
37320 4
};
\addplot [orange, mark=*, mark size=0.7, mark options={solid}, only marks, forget plot]
table {%
35400 5
35430 5
35460 5
35490 5
35520 5
35550 5
35580 5
35610 5
35640 5
35670 5
35790 5
35880 5
35940 5
35970 5
36030 5
36060 5
36090 5
36120 5
36180 5
36270 5
36330 5
36360 5
36390 5
36420 5
36450 5
36480 5
36510 5
36540 5
36570 5
36600 5
36630 5
36660 5
36690 5
36780 5
36810 5
36840 5
36870 5
36900 5
36930 5
36960 5
36990 5
37020 5
37050 5
37080 5
37110 5
37140 5
37170 5
37200 5
37230 5
37260 5
37290 5
37320 5
};
\addplot [semithick, black!30!green, mark=triangle, mark size=1.25, mark options={rotate=180}, only marks, forget plot]
table {%
36390 5
36420 5
36450 5
36480 5
36510 5
36540 5
36570 5
36600 5
36630 5
36660 5
36690 5
36780 5
36810 5
36840 5
36870 5
36900 5
36930 5
36960 5
36990 5
37020 5
37050 5
37080 5
37110 5
37140 5
37170 5
37200 5
37230 5
37260 5
37290 5
37320 5
};
\addplot [orange, mark=*, mark size=0.7, mark options={solid}, only marks, forget plot]
table {%
35400 6
35430 6
35460 6
35490 6
35520 6
35550 6
35580 6
35610 6
35640 6
35670 6
35700 6
35760 6
35940 6
35970 6
36000 6
36030 6
36060 6
36090 6
36120 6
36150 6
36180 6
36270 6
36330 6
36360 6
36390 6
36420 6
36450 6
36480 6
36510 6
36540 6
36570 6
36600 6
36630 6
36690 6
36720 6
36750 6
36780 6
36810 6
36840 6
36870 6
36900 6
36930 6
36960 6
36990 6
37020 6
37050 6
37080 6
37110 6
37140 6
37170 6
37200 6
37230 6
37290 6
37320 6
};
\addplot [semithick, black!30!green, mark=triangle, mark size=1.25, mark options={rotate=180}, only marks, forget plot]
table {%
35400 6
35430 6
35460 6
35490 6
35520 6
35550 6
35580 6
35610 6
35640 6
35670 6
35700 6
35760 6
35940 6
35970 6
36000 6
36030 6
36060 6
36090 6
36120 6
36150 6
36180 6
36270 6
36330 6
36360 6
36390 6
36420 6
36450 6
36480 6
36510 6
36540 6
36570 6
36600 6
36630 6
36690 6
36720 6
36750 6
36780 6
36810 6
36840 6
36870 6
36900 6
36930 6
36960 6
36990 6
37020 6
37050 6
37080 6
37110 6
37140 6
37170 6
37200 6
37230 6
37290 6
};
\addplot [orange, mark=*, mark size=0.7, mark options={solid}, only marks, forget plot]
table {%
35400 7
35430 7
35490 7
35550 7
35580 7
35670 7
35700 7
35910 7
35940 7
36090 7
36720 7
36750 7
36870 7
36900 7
37020 7
37260 7
37290 7
};
\addplot [semithick, black!30!green, mark=triangle, mark size=1.25, mark options={rotate=180}, only marks, forget plot]
table {%
35400 7
35430 7
35490 7
35550 7
35580 7
35670 7
35700 7
35910 7
35940 7
36090 7
36720 7
36750 7
36870 7
36900 7
37020 7
37260 7
37290 7
};
\end{axis}

\end{tikzpicture}}}

  \caption{Hard Urban scenario recording of a walk in the European District of Brussels on December 03, 2023, from 09:50:00 to 10:22:30 UTC.}
\end{figure}
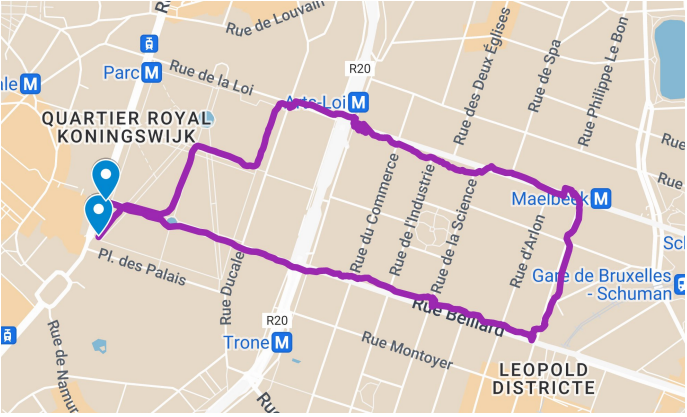
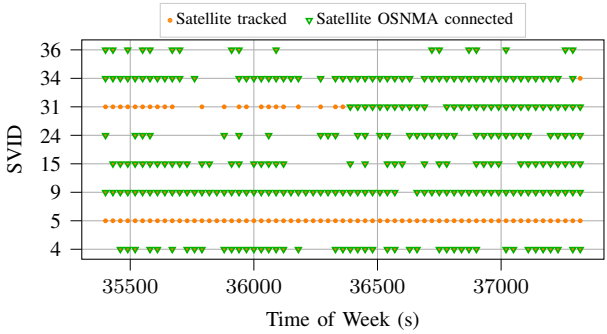

This scenario is a walk in the European District of Brussels, Belgium, on December 03, 2023, from 09:50:00 to 10:22:30 UTC; or GST 1267 35400 to 1267 37350. The trajectory (Fig. \ref{fig:map_eu_district}) starts at the Parc de Bruxelles and quickly heads to the urban canyon of Rue Belliart, Rue de Trèves, and Rue de la Loi. Finally, it returns to the park and ends close to the start location.

During the trajectory, the receiver got navigation data from 8 different satellites (Fig. \ref{fig:satellites_eu_district}). Only SVID 5 did not transmit OSNMA during the scenario; SVID 31 was initially disconnected but started transmitting OSNMA at half the scenario duration. The tracking is generally very volatile, as it corresponds with a hard urban scenario, with several entirely lost sub-frames.

\subsection{Soft Urban Scenario: Brussels, Atomium and Laeken Park}

\begin{figure}
  
  \subfloat[Trajectory followed.\label{fig:map_atomium}]{\centerline{\includegraphics[width=0.5\textwidth,trim={3cm 9cm 2cm 9cm},clip]{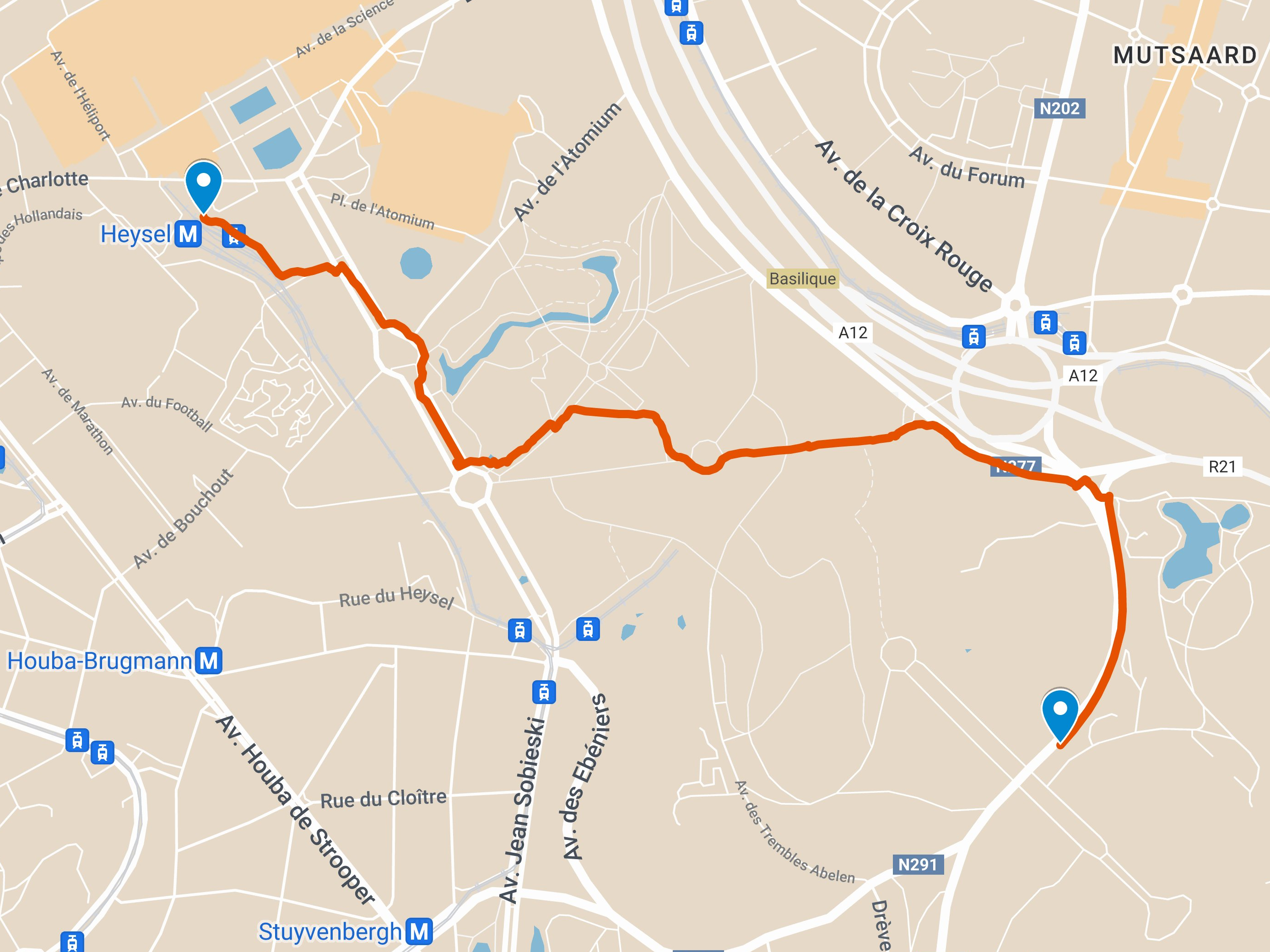}}}\qquad 
  \vspace{0.2cm}
  \subfloat[Galileo satellites tracked and OSNMA connected.\label{fig:satellites_atomium}]{\centerline{\input{media/scenarios/walk_atomium_satellites_scenario}}}

  \caption{Soft Urban scenario recording of a walk around the Atomium of Brussels on December 03, 2023, from 11:03:24 to 11:43:53 UTC.}
\end{figure}

This scenario is a walk around the Atomium and surrounding parks in Brussels, Belgium, on December 03, 2023, from 11:03:24 to 11:43:53 UTC; or GST 1267 39804 to 1267 42233. The trajectory (Fig. \ref{fig:map_atomium}) walks close to the Atomium, enters Osseghem Park, and finally surrounds Laeken Park.

The receiver got navigation data from 9 satellites during the trajectory (Fig. \ref{fig:satellites_atomium}). The number of satellites connected and disconnected is very balanced during the whole scenario, although there is a lot of change in which specific satellites transmit OSNMA.

\subsection{Open-Sky: Leuven, Septentrio Offices}

This scenario is a static recording of 60 minutes from the Septentrio Offices in Leuven, Belgium, on December 20, 2023, from 15:00:00 to 16:00:00 UTC; or GST 1269 313200 to 1269 316800.

\begin{figure}
  \centerline{\input{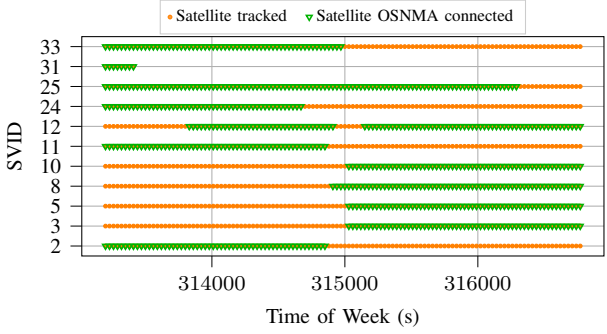}}
  \caption{Galileo satellites tracked and OSNMA connected in the Open-Sky static recording of 60 minutes from the Septentrio offices in Leuven, Belgium, on December 20, 2023; from 15:00:00 to 16:00:00 UTC.}
  \label{fig:satellites_septentrio}
\end{figure}

The satellite visibility of this recording is excellent, as expected in an open-sky situation. A total of 11 satellites are received during the scenario, although the SVID 31 moves under the tracking horizon a few minutes in the recording (Fig. \ref{fig:satellites_septentrio}). All satellites move from connected to disconnected and vice-versa during the recording, but there are always at least four disconnected.

\subsection{Test Vectors: Configuration 2}

The OSNMA receiver guidelines \cite{OSNMA_guidelines} contain several test vectors to validate the implementation of the OSNMA protocol. The test vector titled 'Configuration 2' contains OSNMA data with the same structure as the operational live data described in Fig. \ref{fig:ref_structure}, so it is helpful to test and compare the optimizations. We have run the first 30 minutes of this test vector, simulating from July 26 at 23:59:43 to July 27 at 00:29:43 UTC, or GST 1248 345601 to 1248 347401. These test vectors must be formatted correctly to run in OSNMAlib because they are not chronologically sorted in their original format.

\begin{figure}
  \centerline{\input{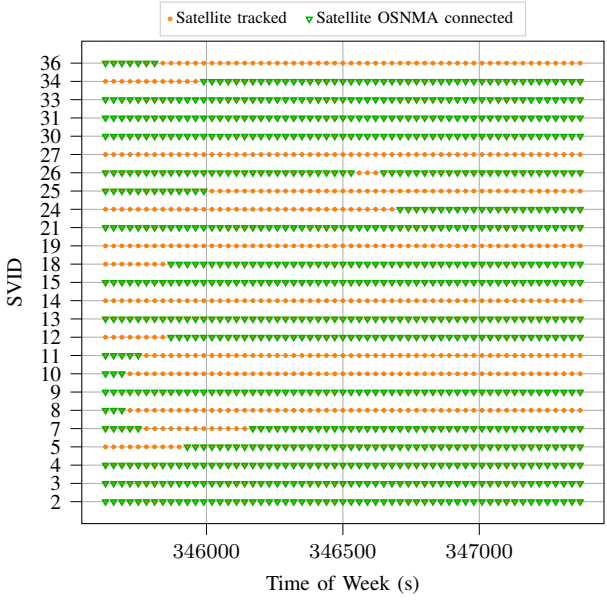}}
  \caption{Galileo satellites tracked and OSNMA connected in Configuration 2 of the test vectors from the OSNMA receiver guidelines \cite{OSNMA_guidelines}. It is a synthetic scenario with all Galileo satellites visible.}
  \label{fig:satellites_config2}
\end{figure}

A particular characteristic of the test vectors is that they contain data from 25 Galileo satellites, which is impossible in a live recording (see Fig. \ref{fig:satellites_config2}). Moreover, they emulate a perfect reception with no pages lost. Therefore, while we cannot directly extrapolate the results to a real scenario, they are useful to validate if the tag-data link optimizations work.

\begin{figure*}[t]
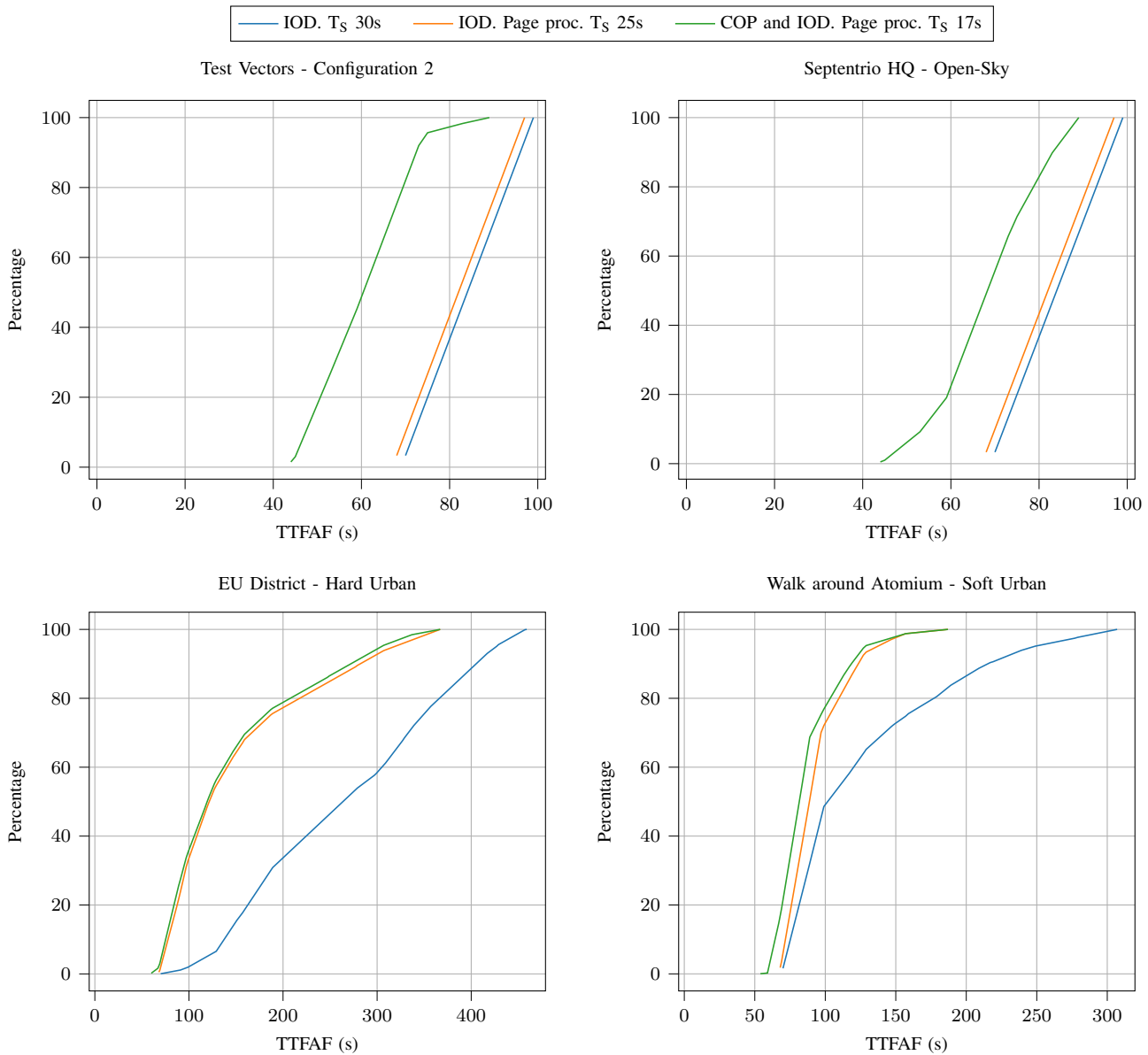

\centering

    \begin{tikzpicture}
    
    \definecolor{darkgray176}{RGB}{176,176,176}
    \definecolor{darkorange25512714}{RGB}{255,127,14}
    \definecolor{forestgreen4416044}{RGB}{44,160,44}
    \definecolor{lightgray204}{RGB}{204,204,204}
    \definecolor{steelblue31119180}{RGB}{31,119,180}
    
    \begin{groupplot}[
        group style={
            group size=2 by 2,
            vertical sep=2cm,
            horizontal sep=2cm,
        },
        legend cell align={left},
        legend style={
            legend columns=3,
            /tikz/every even column/.append style={column sep=0.4cm}
        }
      ]
        \nextgroupplot[
            tick align=outside,
            tick pos=left,
            title={Test Vectors - Configuration 2},
            x grid style={darkgray176},
            xlabel={TTFAF (s)},
            xmajorgrids,
            xmin=-1.75, xmax=101.75,
            xtick style={color=black},
            y grid style={darkgray176},
            ylabel={Percentage},
            ymajorgrids,
            ymin=-3.425, ymax=104.925,
            ytick style={color=black},
            legend to name={CDFCommonLegend},
            ]
        \input{media/results_one_figure/config2_cdf}
        \addlegendimage{semithick, steelblue31119180}
        \addlegendentry{IOD. T\textsubscript{S} 30s}
        \addlegendimage{semithick, darkorange25512714}
        \addlegendentry{IOD. Page proc. T\textsubscript{S} 25s}
        \addlegendimage{semithick, forestgreen4416044}
        \addlegendentry{COP and IOD. Page proc. T\textsubscript{S} 17s}
    
        \nextgroupplot[
            tick align=outside,
            tick pos=left,
            title={Septentrio HQ - Open-Sky},
            x grid style={darkgray176},
            xlabel={TTFAF (s)},
            xmajorgrids,
            xmin=-1.75, xmax=101.75,
            xtick style={color=black},
            y grid style={darkgray176},
            ylabel={Percentage},
            ymajorgrids,
            ymin=-4.475, ymax=104.975,
            ytick style={color=black},
            ]
        \input{media/results_one_figure/open_sky_cdf}
    
        \nextgroupplot[
            tick align=outside,
            tick pos=left,
            title={EU District - Hard Urban},
            x grid style={darkgray176},
            xlabel={TTFAF (s)},
            xmajorgrids,
            xmin=-6.0, xmax=478.95,
            xtick style={color=black},
            y grid style={darkgray176},
            ylabel={Percentage},
            ymajorgrids,
            ymin=-4.94615384615385, ymax=104.997435897436,
            ytick style={color=black},
            ]
        \input{media/results_one_figure/park_eu_cdf}
    
        \nextgroupplot[
            tick align=outside,
            tick pos=left,
            title={Walk around Atomium - Soft Urban},
            x grid style={darkgray176},
            xlabel={TTFAF (s)},
            xmajorgrids,
            xmin=-4.0, xmax=319.65,
            xtick style={color=black},
            y grid style={darkgray176},
            ylabel={Percentage},
            ymajorgrids,
            ymin=-4.95677233429395, ymax=104.997941539728,
            ytick style={color=black},
            ]
        \input{media/results_one_figure/walk_atomium_cdf}
        
    \end{groupplot}
    \node (legend) at ($(group c1r1.center)!0.5!(group c2r1.center)+(0,4cm)$) {\pgfplotslegendfromname{CDFCommonLegend}};
    \end{tikzpicture}

\caption{The page-level processing optimization improves the TTFAF on the scenarios where pages are lost, such as the urban scenarios. The COP-IOD optimization improves as expected the TTFAF only in the scenarios where a lot of satellites are visible, while it struggles to bring any benefit in the urban scenarios.}
\label{fig:TTFAF_cdf_one_plot}
\end{figure*}

\begin{figure*}[t]
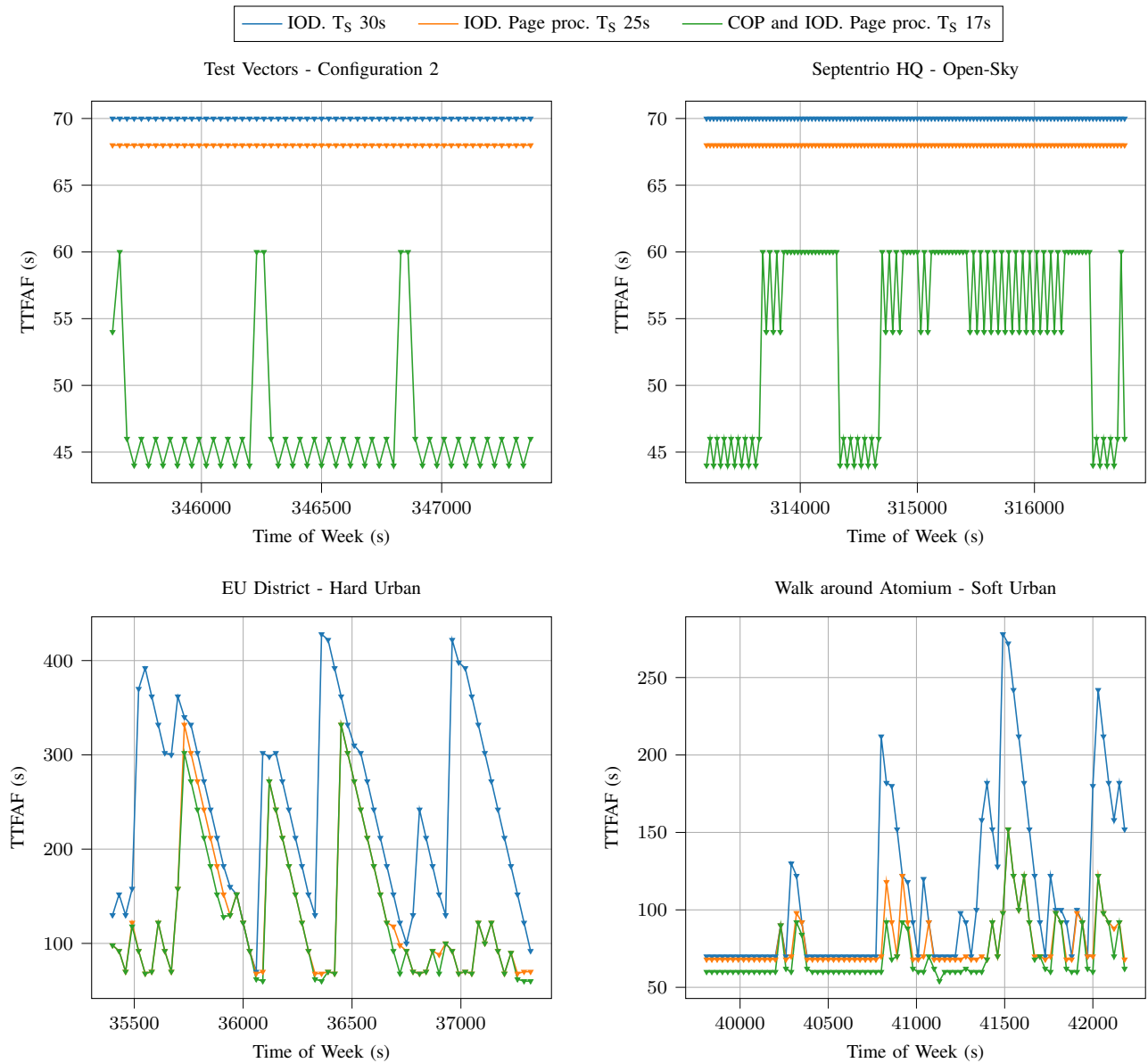

\centering

    \begin{tikzpicture}
    
    \definecolor{darkgray176}{RGB}{176,176,176}
    \definecolor{darkorange25512714}{RGB}{255,127,14}
    \definecolor{forestgreen4416044}{RGB}{44,160,44}
    \definecolor{lightgray204}{RGB}{204,204,204}
    \definecolor{steelblue31119180}{RGB}{31,119,180}
    
    \begin{groupplot}[
        group style={
            group size=2 by 2,
            vertical sep=2cm,
            horizontal sep=2cm,
        },
        legend cell align={left},
        legend style={
            legend columns=3,
            /tikz/every even column/.append style={column sep=0.4cm}
        }
      ]
        \nextgroupplot[
            tick align=outside,
            tick pos=left,
            title={Test Vectors - Configuration 2},
            x grid style={darkgray176},
            xmajorgrids,
            xmin=345543, xmax=347457,
            xtick style={color=black},
            y grid style={darkgray176},
            ylabel={TTFAF (s)},
            xlabel={Time of Week (s)},
            ymajorgrids,
            ymin=42.7, ymax=71.3,
            ytick style={color=black},
            legend to name={TTFAFCommonLegend},
            ]
        \input{media/results_one_figure/config2_ttfaf}
        \addlegendimage{semithick, steelblue31119180}
        \addlegendentry{IOD. T\textsubscript{S} 30s}
        \addlegendimage{semithick, darkorange25512714}
        \addlegendentry{IOD. Page proc. T\textsubscript{S} 25s}
        \addlegendimage{semithick, forestgreen4416044}
        \addlegendentry{COP and IOD. Page proc. T\textsubscript{S} 17s}
    
        \nextgroupplot[
            tick align=outside,
            tick pos=left,
            title={Septentrio HQ - Open-Sky},
            x grid style={darkgray176},
            xmajorgrids,
            xmin=313021.5, xmax=316948.5,
            xtick style={color=black},
            y grid style={darkgray176},
            ylabel={TTFAF (s)},
            xlabel={Time of Week (s)},
            ymajorgrids,
            ymin=42.7, ymax=71.3,
            ytick style={color=black},
            ]
        \input{media/results_one_figure/open_sky_ttfaf}
    
        \nextgroupplot[
            tick align=outside,
            tick pos=left,
            title={EU District - Hard Urban},
            x grid style={darkgray176},
            xmajorgrids,
            xmin=35304, xmax=37416,
            xtick style={color=black},
            y grid style={darkgray176},
            ylabel={TTFAF (s)},
            xlabel={Time of Week (s)},
            ymajorgrids,
            ymin=41.6, ymax=446.4,
            ytick style={color=black},
            ]
        \input{media/results_one_figure/park_eu_ttfaf}
    
        \nextgroupplot[
            tick align=outside,
            tick pos=left,
            title={Walk around Atomium - Soft Urban},
            x grid style={darkgray176},
            xmajorgrids,
            xmin=39691.5, xmax=42298.5,
            xtick style={color=black},
            y grid style={darkgray176},
            ylabel={TTFAF (s)},
            xlabel={Time of Week (s)},
            ymajorgrids,
            ymin=42.8, ymax=289.2,
            ytick style={color=black},
            ]
        \input{media/results_one_figure/walk_atomium_ttfaf}
        
    \end{groupplot}
    \node (legend) at ($(group c1r1.center)!0.5!(group c2r1.center)+(0,4cm)$) {\pgfplotslegendfromname{TTFAFCommonLegend}};
    \end{tikzpicture}

\caption{The figures display the minimum TTFAF value obtained on each sub-frame. The theoretical minimum value with the COP-IOD optimization is 44 or 46 seconds. This value is only consistently achieved in the test vectors (except for the sub-frames with change of navigation data) and in some sub-frames of the Open-Sky scenario. The COP-IOD minimum value is never reached in the urban scenarios due to the high number of lost pages. However, for the same reason, the page-level processing optimization substantially improves the TTFAF values in the urban scenarios.}
\label{fig:min_TTFAF_one_plot}
\end{figure*}

\section{TEST RESULTS}

We implemented the optimizations in OSNMAlib in a flexible way so that they can be turned on or off at choice. To obtain multiple TTFAF values from the continuous recordings, we replayed the logs in OSNMAlib but started to process them each time one second later. With this technique, we can emulate a receiver powering up at any moment of the recording and obtain all the TTFAF values needed to evaluate the optimizations. Therefore, the number of data points is directly the number of seconds on each scenario.

We decided to group the described optimizations into three accumulative groups to visualize their effects easily:
\begin{itemize}
    \item Standard OSNMA: Uses the IOD optimization to regenerate navigation data and the default T\textsubscript{S} set to T\textsubscript{L} (30 seconds). While a standard OSNMA may not include the IOD optimization, it is briefly described in the OSNMA ICD, was present in the first version of OSNMAlib, and is already used in other state-of-the-art implementations. Hence, we use this configuration as a baseline.
    \item Page-Level Processing and Tighter Time Synchronization: Uses the IOD optimization, a T\textsubscript{S} of 25 seconds to use the IOD optimization at its full potential, and the page-level processing technique to extract valid navigation data from broken sub-frames.
    \item COP and IOD, with Page-Level Processing and Tighter Time Synchronization: Uses the COP-IOD optimization to regenerate and propagate navigation data, a T\textsubscript{S} of 17 seconds to use completely the COP optimization, and page-level processing. 
\end{itemize}

The results are presented in a cumulative distribution function (CDF) for each scenario to provide a global view of the optimization performance (Fig. \ref{fig:TTFAF_cdf_one_plot}). Additionally, in Fig. \ref{fig:min_TTFAF_one_plot} we present the minimum TTFAF value obtained in each sub-frame to evaluate how the optimizations improve the TTFAF at different time periods.

\subsection{Page-Level Tag and Key Processing}

The page-level processing optimization works as expected: it improves the TTFAF in cases where Galileo I/NAV pages are lost. The two urban scenarios show a clear improvement between the case with page-level processing and the case without it (Fig. \ref{fig:TTFAF_cdf_one_plot}). Due to the buildings and trees, nearly any satellite has dropped pages at some point, and the optimization extracts all it can from the left pages. For example, in the Hard Urban scenario, nearly 80\% of the TTFAF values are lower than 200 seconds when using page-level processing, while the TTFAF increases to 360 seconds for the case without this optimization. Unsurprisingly, the improvement is more significant in the Hard Urban scenario than in the Soft Urban case, where fewer pages are lost.

When looking at the minimum TTFAF value per sub-frame (Fig. \ref{fig:min_TTFAF_one_plot}) for the same urban scenarios, the effect of the harsh environment is displayed in the form of time spikes. In some cases, the page processing optimization follows the same spike as the not-optimized case but with slightly lower values. However, when this does not happen, the improvement is substantial (for example, around Time of Week 37000 in the Hard Urban scenario).

In the Open-Sky scenario and test vectors, the two-second improvement observed in the minimum TTFAF per sub-frame and in the displacement of the CDF is due to the reduction of the T\textsubscript{S} to 25 seconds and not to the page-level processing. Reducing the T\textsubscript{S} allows linking navigation data of two sub-frames using the IOD of the WT 3 instead of 1. The WT 3 is transmitted 2 seconds after the WT 1, hence the improvement of 2 seconds in the minimum TTFAF when reducing the T\textsubscript{S} to 25 seconds.

The ineffectiveness of the page-level processing for the test vectors is expected: the synthetic nature of the scenario implies that no pages are lost. In the open-sky scenario we recorded no satellite loses relevant pages for OSNMA, which is a possible situation. Nevertheless, note that this can differ for other open-sky scenarios: some low-elevation satellites might lose pages.

\subsection{COP-IOD Tag-Data Link}

The COP-IOD tag-data link optimization struggles to yield any improvement in the urban scenarios (Fig. \ref{fig:TTFAF_cdf_one_plot}). The essential requirement of obtaining two tags for the same satellite and navigation data in two consecutive sub-frames is hardly met due to the fading characteristic of these environments. Also, the reduced number of satellites in view makes this requirement even harder to fulfill. Still, it improves slightly more in the Soft Urban than in the Hard Urban scenario.

However, the optimization works according to the theory in the test vectors, improving the TTFAF very substantially. Some cases worse than expected can be seen as sub-frames with a minimum TTFAF of 60 seconds in Fig. \ref{fig:min_TTFAF_one_plot} because the test vectors contain sub-frames with change of navigation data. When sufficient navigation data changes, the optimization cannot make assumptions based on the COP value for all tags, degrading the TTFAF. Despite that, it is always better than the cases with only the IOD optimization. Moreover, we can see how the lowest case for each sub-frame alternates between 44 and 46 seconds, determined by whether the tag sequence is for the odd or even sub-frame (see in Fig. \ref{fig:ref_structure} the position of the last cross-authentication tag {\tt E00}).

Strangely, in the Open-Sky scenario, the COP-IOD optimization does not seem to work as well as theorized, even when tracking 10 satellites for most of the time. The results are good; the improvement, when compared with the IOD optimization only values, is clear and huge, but we are in several sub-frames far away from the 44 to 46 seconds mark. Additionally, we can see how the minimum TTFAF value for the sub-frames is discrete: 60, 54, 46, and 44 seconds. These values are directly linked to the position of the ADKD0 tags in the tag sequence, described in Fig. \ref{fig:ref_structure}.

When a receiver implementing the COP-IOD optimization starts aligned with the beginning of the sub-frame, it receives four ADKD0 tags on that sub-frame from each connected satellite. If four of these tags are repeated in the next sub-frame, a TTFAF of 60 seconds can be obtained. This case is very likely with 10 satellites in view for the Open-Sky scenario. Thus, we do not see any sub-frame with a minimum TTFAF greater than 60 seconds.

If the receiver starts later within the sub-frame and misses the first tag, it also loses the ability to authenticate the flex tag positions, effectively losing all flex tags. Therefore, it can only use the ADKD0 tags indicated with \texttt{00E} in Fig. \ref{fig:ref_structure}. The discrete TTFAF values for the Open-Sky scenario in Fig. \ref{fig:min_TTFAF_one_plot} are obtained when the receiver starts just before this ADKD0 tags.

For each of the three tested optimization combinations, we have chosen the lowest, average, and percentile 95 values as relevant TTFAF metrics and displayed them in Tables \ref{tab:ttfaf_iod_metrics}, \ref{tab:ttfaf_iod_page_metrics}, and \ref{tab:ttfaf_cop_iod_page_metrics}. The combination of IOD and Cut-Off Point tag-data link with the page-level processing and a T\textsubscript{S} of 17 seconds always brings the best results regardless of the circumstance.

\begin{table}[h]
\renewcommand{\arraystretch}{1.3}
\caption{TTFAF metrics using the IOD Data Link optimization with a T\textsubscript{S} of 30 seconds}
\begin{center}
\begin{tabular}{|l|c|c|c|}
\hline
\textbf{Optimization}   & \textbf{Lowest (s)} & \textbf{Average (s)} & \textbf{P95 (s)} \\ \hline
Test Vectors            &  70.0 & 84.5  & 98.0  \\ \hline
Open-Sky                &  70.0 & 84.5  & 98.0  \\ \hline
Soft Urban              &  70.0 & 127.5 & 248.0 \\ \hline
Hard Urban              &  70.0 & 266.1 & 427.0 \\ \hline
\end{tabular}
\label{tab:ttfaf_iod_metrics}
\end{center}
\end{table}

\begin{table}[h]
\renewcommand{\arraystretch}{1.3}
\caption{TTFAF metrics using the IOD Data Link optimization with a T\textsubscript{S} of 25 seconds and page-level processing}
\begin{center}
\begin{tabular}{|l|c|c|c|}
\hline
\textbf{Optimization}   & \textbf{Lowest (s)} & \textbf{Average (s)} & \textbf{P95 (s)} \\ \hline
Test Vectors            &  68.0 & 82.5  & 96.0  \\ \hline
Open-Sky                &  68.0 & 82.5  & 96.0  \\ \hline
Soft Urban              &  68.0 & 94.1  & 137.0 \\ \hline
Hard Urban              &  68.0 & 151.1 & 318.5 \\ \hline
\end{tabular}
\label{tab:ttfaf_iod_page_metrics}
\end{center}
\end{table}

\begin{table}[h]
\renewcommand{\arraystretch}{1.3}
\caption{TTFAF metrics using the COP-IOD Tag-Data Link optimization with a T\textsubscript{S} of 17 seconds and page-level processing}
\begin{center}
\begin{tabular}{|l|c|c|c|}
\hline
\textbf{Optimization}   & \textbf{Lowest (s)} & \textbf{Average (s)} & \textbf{P95 (s)} \\ \hline
Test Vectors            &  44.0 & 60.9  & 75.0  \\ \hline
Open-Sky                &  44.0 & 68.8  & 87.0  \\ \hline
Soft Urban              &  54.0 & 87.5  & 129.0 \\ \hline
Hard Urban              &  60.0 & 146.1 & 305.0 \\ \hline
\end{tabular}
\label{tab:ttfaf_cop_iod_page_metrics}
\end{center}
\end{table}

\subsection{OSNMA Cross-Authentication Algorithm}

The reason why, even in an open-sky scenario, the COP optimization is not working as well as expected lies in the OSNMA cross-authentication algorithm. Currently, OSNMA only transmits cross-authentication tags for disconnected satellites, and this behavior creates an imbalance in the number of ADKD0 tags a satellite receives during a sub-frame conditioned by its connection status.

If a satellite is connected, it will only receive one ADKD0 tag for the whole sub-frame: the self-authenticating tag, which is always transmitted at the first position of the tag sequence (see {\tt 00S} in Fig. \ref{fig:ref_structure}). However, if the satellite is disconnected, it will get multiple cross-authenticating ADKD0 tags from connected satellites. These tags are transmitted in the cross-authentication positions, which are currently three per sub-frame (see {\tt 00E} and {\tt FLX} in Fig. \ref{fig:ref_structure}).

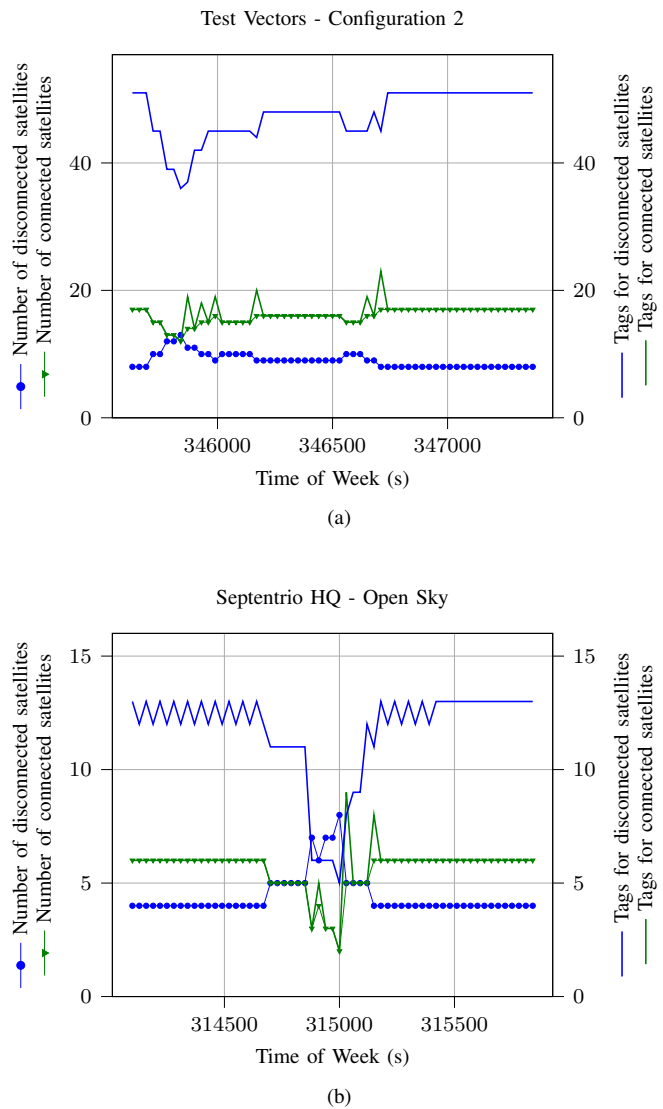
\begin{figure}[!t]
  \centering
  \subfloat[\label{fig:config2_tags}]{
\begin{tikzpicture}

\definecolor{darkgray176}{RGB}{176,176,176}
\definecolor{green01270}{RGB}{0,127,0}
\definecolor{lightgray204}{RGB}{204,204,204}

\begin{axis}[
tick align=outside,
tick pos=left,
title={Test Vectors - Configuration 2},
width=.41\textwidth,
x grid style={darkgray176},
xmajorgrids,
xmin=345543, xmax=347457,
xtick style={color=black},
xlabel={Time of Week (s)},
y grid style={darkgray176},
ylabel style={align=center},
ylabel={\ref{config2_tracked_sats} Number of disconnected satellites\\ \ref{config2_OSNMA_tracked_sats} Number of connected satellites},
ymajorgrids,
ymin=0, ymax=57,
ytick style={color=black}
]

\addlegendimage{very thin, blue, mark=*, mark size=1.5, mark options={solid}}
\label{config2_tracked_sats}

\addlegendimage{very thin, green01270, mark=triangle*, mark size=1.5, mark options={solid,rotate=180}}
\label{config2_OSNMA_tracked_sats}

\addplot [very thin, blue, mark=*, mark size=1, mark options={solid}]
table {%
345630 8
345660 8
345690 8
345720 10
345750 10
345780 12
345810 12
345840 13
345870 11
345900 11
345930 10
345960 10
345990 9
346020 10
346050 10
346080 10
346110 10
346140 10
346170 9
346200 9
346230 9
346260 9
346290 9
346320 9
346350 9
346380 9
346410 9
346440 9
346470 9
346500 9
346530 9
346560 10
346590 10
346620 10
346650 9
346680 9
346710 8
346740 8
346770 8
346800 8
346830 8
346860 8
346890 8
346920 8
346950 8
346980 8
347010 8
347040 8
347070 8
347100 8
347130 8
347160 8
347190 8
347220 8
347250 8
347280 8
347310 8
347340 8
347370 8
};
\addplot [very thin, green01270, mark=triangle*, mark size=1, mark options={solid,rotate=180}]
table {%
345630 17
345660 17
345690 17
345720 15
345750 15
345780 13
345810 13
345840 12
345870 14
345900 14
345930 15
345960 15
345990 16
346020 15
346050 15
346080 15
346110 15
346140 15
346170 16
346200 16
346230 16
346260 16
346290 16
346320 16
346350 16
346380 16
346410 16
346440 16
346470 16
346500 16
346530 16
346560 15
346590 15
346620 15
346650 16
346680 16
346710 17
346740 17
346770 17
346800 17
346830 17
346860 17
346890 17
346920 17
346950 17
346980 17
347010 17
347040 17
347070 17
347100 17
347130 17
347160 17
347190 17
347220 17
347250 17
347280 17
347310 17
347340 17
347370 17
};
\end{axis}

\begin{axis}[
width=.41\textwidth,
tick align=outside,
xtick style={draw=none},
xmin=345543, xmax=347457,
xticklabels={},
y grid style={darkgray176},
ylabel style={align=center},
ylabel={\ref{conf2_tags_not_connected} Tags for disconnected satellites\\ \ref{conf2_tags_connected} Tags for connected satellites},
ymin=0, ymax=57,
ytick pos=right,
ytick style={color=black},
yticklabel style={anchor=west}
]
\addplot [semithick, green01270]
table {%
345630 17
345660 17
345690 17
345720 15
345750 15
345780 13
345810 13
345840 12
345870 19
345900 14
345930 18
345960 15
345990 19
346020 15
346050 15
346080 15
346110 15
346140 15
346170 20
346200 16
346230 16
346260 16
346290 16
346320 16
346350 16
346380 16
346410 16
346440 16
346470 16
346500 16
346530 16
346560 15
346590 15
346620 15
346650 19
346680 16
346710 23
346740 17
346770 17
346800 17
346830 17
346860 17
346890 17
346920 17
346950 17
346980 17
347010 17
347040 17
347070 17
347100 17
347130 17
347160 17
347190 17
347220 17
347250 17
347280 17
347310 17
347340 17
347370 17
};\label{conf2_tags_connected}
\addplot [semithick, blue]
table {%
345630 51
345660 51
345690 51
345720 45
345750 45
345780 39
345810 39
345840 36
345870 37
345900 42
345930 42
345960 45
345990 45
346020 45
346050 45
346080 45
346110 45
346140 45
346170 44
346200 48
346230 48
346260 48
346290 48
346320 48
346350 48
346380 48
346410 48
346440 48
346470 48
346500 48
346530 48
346560 45
346590 45
346620 45
346650 45
346680 48
346710 45
346740 51
346770 51
346800 51
346830 51
346860 51
346890 51
346920 51
346950 51
346980 51
347010 51
347040 51
347070 51
347100 51
347130 51
347160 51
347190 51
347220 51
347250 51
347280 51
347310 51
347340 51
347370 51
};\label{conf2_tags_not_connected}
\end{axis}

\end{tikzpicture}}
  \vspace{0.3cm}
  \subfloat[\label{fig:open_sky_tags}]{
\begin{tikzpicture}

\definecolor{darkgray176}{RGB}{176,176,176}
\definecolor{green01270}{RGB}{0,127,0}
\definecolor{lightgray204}{RGB}{204,204,204}

\begin{axis}[
width=.41\textwidth,
tick align=outside,
tick pos=left,
title={Septentrio HQ - Open Sky},
x grid style={darkgray176},
xmajorgrids,
xmin=314013, xmax=315927,
xtick style={color=black},
xlabel={Time of Week (s)},
y grid style={darkgray176},
ylabel style={align=center},
ylabel={\ref{tracked_sats} Number of disconnected satellites\\ \ref{OSNMA_tracked_sats} Number of connected satellites},
ymajorgrids,
ymin=0, ymax=16,
ytick style={color=black}
]

\addlegendimage{very thin, blue, mark=*, mark size=1.5, mark options={solid}}
\label{tracked_sats}

\addlegendimage{very thin, green01270, mark=triangle*, mark size=1.5, mark options={solid,rotate=180}}
\label{OSNMA_tracked_sats}

\addplot [very thin, blue, mark=*, mark size=1, mark options={solid}]
table {%
314100 4
314130 4
314160 4
314190 4
314220 4
314250 4
314280 4
314310 4
314340 4
314370 4
314400 4
314430 4
314460 4
314490 4
314520 4
314550 4
314580 4
314610 4
314640 4
314670 4
314700 5
314730 5
314760 5
314790 5
314820 5
314850 5
314880 7
314910 6
314940 7
314970 7
315000 8
315030 5
315060 5
315090 5
315120 5
315150 4
315180 4
315210 4
315240 4
315270 4
315300 4
315330 4
315360 4
315390 4
315420 4
315450 4
315480 4
315510 4
315540 4
315570 4
315600 4
315630 4
315660 4
315690 4
315720 4
315750 4
315780 4
315810 4
315840 4
}; 
\addplot [very thin, green01270, mark=triangle*, mark size=1, mark options={solid,rotate=180}]
table {%
314100 6
314130 6
314160 6
314190 6
314220 6
314250 6
314280 6
314310 6
314340 6
314370 6
314400 6
314430 6
314460 6
314490 6
314520 6
314550 6
314580 6
314610 6
314640 6
314670 6
314700 5
314730 5
314760 5
314790 5
314820 5
314850 5
314880 3
314910 4
314940 3
314970 3
315000 2
315030 5
315060 5
315090 5
315120 5
315150 6
315180 6
315210 6
315240 6
315270 6
315300 6
315330 6
315360 6
315390 6
315420 6
315450 6
315480 6
315510 6
315540 6
315570 6
315600 6
315630 6
315660 6
315690 6
315720 6
315750 6
315780 6
315810 6
315840 6
};
\end{axis}

\begin{axis}[
width=.41\textwidth,
tick align=outside,
xtick style={draw=none},
xmin=314013, xmax=315927,
xticklabels={},
y grid style={darkgray176},
ylabel style={align=center},
ylabel={\ref{tags_not_connected} Tags for disconnected satellites\\ \ref{tags_connected} Tags for connected satellites},
ymin=0, ymax=16,
ytick pos=right,
ytick style={color=black},
yticklabel style={anchor=west}
]
\addplot [semithick, green01270]
table {%
314100 6
314130 6
314160 6
314190 6
314220 6
314250 6
314280 6
314310 6
314340 6
314370 6
314400 6
314430 6
314460 6
314490 6
314520 6
314550 6
314580 6
314610 6
314640 6
314670 6
314700 5
314730 5
314760 5
314790 5
314820 5
314850 5
314880 3
314910 5
314940 3
314970 3
315000 2
315030 9
315060 5
315090 5
315120 5
315150 8
315180 6
315210 6
315240 6
315270 6
315300 6
315330 6
315360 6
315390 6
315420 6
315450 6
315480 6
315510 6
315540 6
315570 6
315600 6
315630 6
315660 6
315690 6
315720 6
315750 6
315780 6
315810 6
315840 6
};\label{tags_connected}
\addplot [semithick, blue]
table {%
314100 13
314130 12
314160 13
314190 12
314220 13
314250 12
314280 13
314310 12
314340 13
314370 12
314400 13
314430 12
314460 13
314490 12
314520 13
314550 12
314580 13
314610 12
314640 13
314670 12
314700 11
314730 11
314760 11
314790 11
314820 11
314850 11
314880 6
314910 6
314940 6
314970 6
315000 5
315030 8
315060 9
315090 9
315120 12
315150 11
315180 13
315210 12
315240 13
315270 12
315300 13
315330 12
315360 13
315390 12
315420 13
315450 13
315480 13
315510 13
315540 13
315570 13
315600 13
315630 13
315660 13
315690 13
315720 13
315750 13
315780 13
315810 13
315840 13
};\label{tags_not_connected}
\end{axis}

\end{tikzpicture}}

  \caption{On the left y-axis, connected and disconnected satellites. On the right y-axis, the number of authentication tags received per sub-frame for connected and disconnected satellites. A connected satellite only gets 1 tag per sub-frame, while a disconnected satellite gets up to 5 tags per sub-frame on average.}
  \label{fig:tags_both}
\end{figure}

The tag unbalance becomes apparent when examining the number of tags received for connected and disconnected satellites in the test vectors and the Open-Sky scenarios (Fig. \ref{fig:tags_both}). The number of tags per sub-frame for connected satellites is always the same as the number of connected satellites (hence, one tag per satellite). Yet, there are some sub-frames where there is more than one tag per satellite: when a previously disconnected satellite joins the OSNMA transmission. In those cases, because the tags are always transmitted for data in the previous sub-frame, the system still transmits tags for the satellite's data before the satellite starts to transmit OSNMA.

On the other hand, the number of tags received for disconnected satellites is up to 5 times the number of disconnected satellites in the test vectors (Fig. \ref{fig:config2_tags}). The ratio is a bit lower for the Open-Sky scenario (Fig. \ref{fig:open_sky_tags}) because not all satellites are in view, so some tags are lost. In either case, the tags received for disconnected satellites are much more than those for connected satellites.

Another point of discussion is which disconnected satellites are selected for the cross-authentication positions. In the present OSNMA configuration, the connected satellites transmit every sub-frame one tag for the closest and second closest disconnected satellites, and one tag that alternates between the third and fourth closest disconnected satellites \cite{enc_osnmasequence}.

\subsection{Cross-Authentication Algorithm Impact on the COP-IOD Optimization}

\begin{figure}
  \subfloat[With 4 connected satellites in view, they all self-authenticate using the first tag of the sub-frame. The COP-IOD optimization obtains a lowest TTFAF of 60 seconds.\label{fig:cross_auth_4_connected}]{
  \centerline{
    \includegraphics[width=0.30\textwidth]{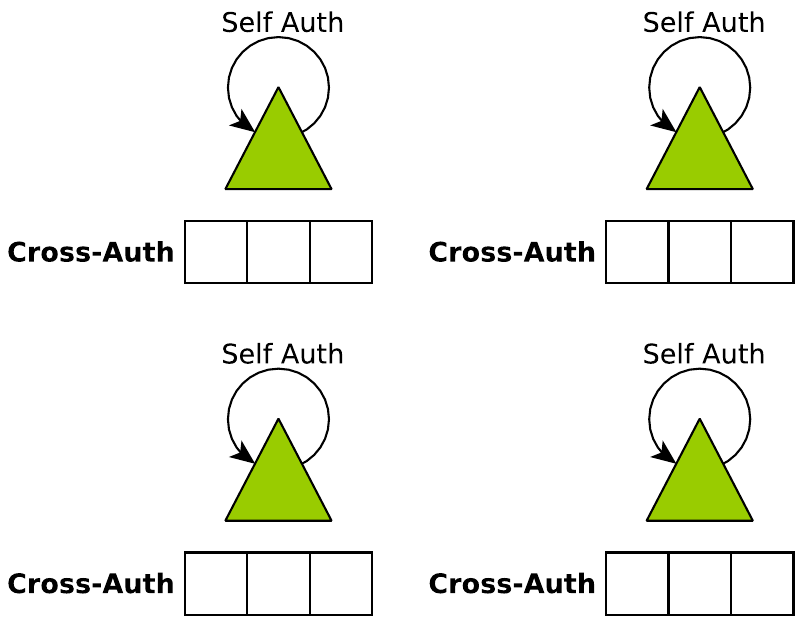}}}\qquad
  \subfloat[With 2 connected and 4 disconnected satellites in view, the connected satellites cross-authenticate the disconnected. These tags are transmitted later in the sub-frame, allowing for a lowest TTFAF of 54 seconds.\label{fig:cross_auth_2_connected_4_disconnected}]{
  \centerline{
    \includegraphics[width=0.35\textwidth]{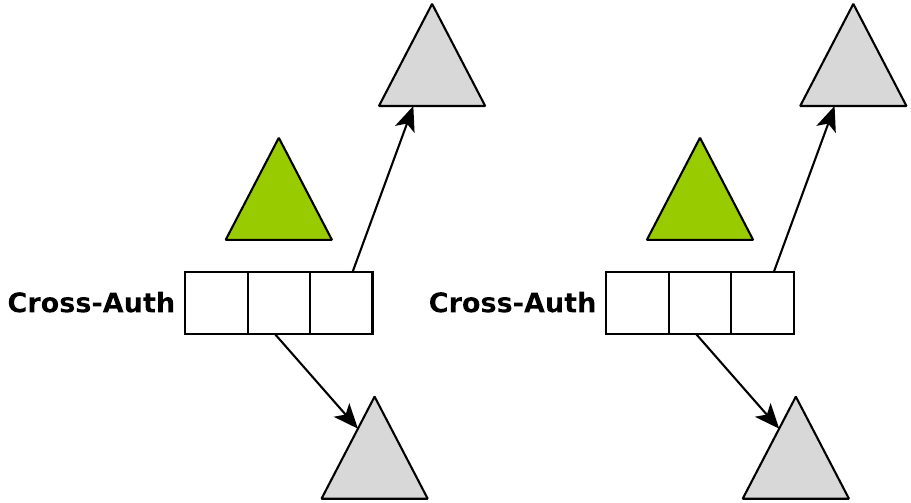}}}\qquad
  \subfloat[With 4 connected and 4 disconnected satellites in view, the connected satellites cross-authenticate the disconnected with a tag in the last position. Hence, the lowest TTFAF can be of 44 seconds.\label{fig:cross_auth_4_connected_4_disconnected}]{
  \centerline{
    \includegraphics[width=0.35\textwidth]{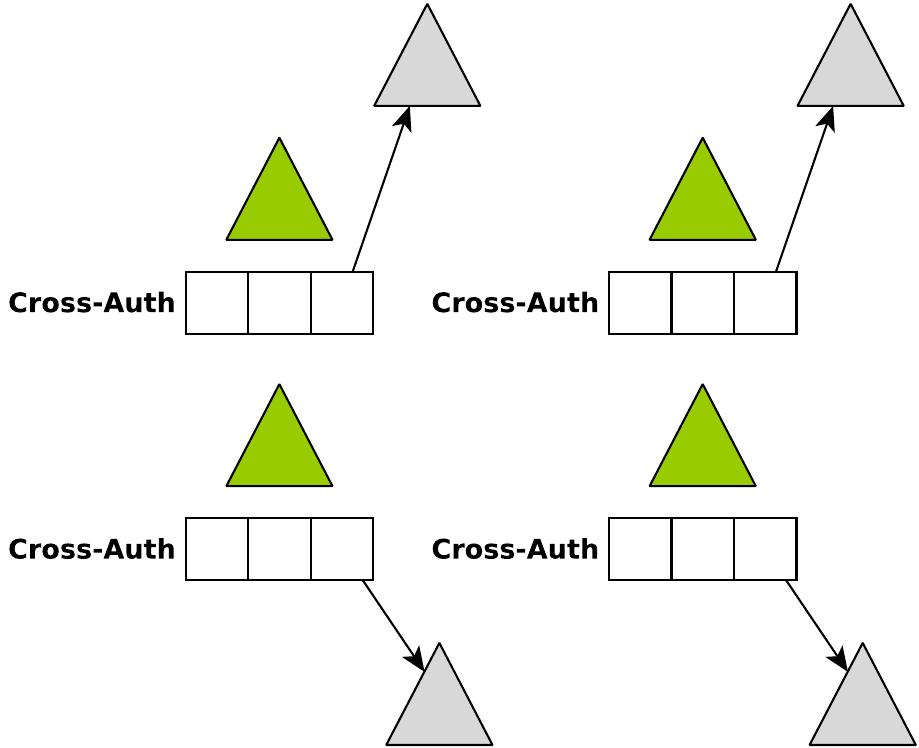}}}
  \vspace{0.3cm}
  \caption{The figure shows the COP-IOD optimization performance on multiple relevant scenarios. The triangle shapes represent satellites in view, with the green color for connected and the gray for disconnected. The arrows indicate for which satellite are the authentication tags issued.}
  \label{fig:cross_auth_figure}
\end{figure}

Cross-authenticating only the disconnected satellites might maximize all-in-view satellite authentication with few connected satellites, but it hampers the performance of the COP-IOD optimization. Moreover, the more satellites become connected, the less tags are transmitted for satellites in view.

For example, in a seemingly good scenario with four connected satellites in view, an OSNMA receiver with the COP-IOD optimization will only be able to achieve a lowest TTFAF of 60 seconds, with an average of 74.5 seconds. This is because the only ADKD0 tag the satellites are getting is transmitted in the first position of the sequence, so if the receiver starts two seconds after the beginning of the sub-frame, it is sure not to receive any tag for that satellite for the rest 28 seconds of the sub-frame (Fig. \ref{fig:cross_auth_4_connected}). However, it will still get better TTFAF values than using only the IOD optimization, with an average of 82.5 seconds, or no tag-data link optimization, with an average of 104.5 seconds.

With six satellites in view and only two of them connected, the COP-IOD optimization obtains better TTFAF values than with 4 connected satellites. Because the cross-authentication tag positions are situated later in the sub-frame, a receiver can start processing later and still receive tags for satellites in view \ref{fig:cross_auth_2_connected_4_disconnected}. In this scenario, and assuming the tags are transmitted in the last two positions, the lowest possible TTFAF is 54 seconds, with an average of 71.5 seconds. However, with the current OSNMA configuration (Fig. \ref{fig:ref_structure}), this scenario can only happen in the odd sub-frames where there are two \texttt{00E} positions. In the even sub-frames, the \texttt{FLX} positions can not be used when the receiver misses the first tag \texttt{00S}.

It is not until we have 8 satellites in view, 4 connected and 4 disconnected, that the COP-IOD optimization works as well as theorized. In this scenario, the 4 connected satellites transmit cross-authentication tags in the last position of the sequence for the 4 disconnected satellites \ref{fig:cross_auth_4_connected_4_disconnected}. Thus, a receiver can start much later in the sub-frame and still receive tags for satellites in view. In this situation, the lowest possible TTFAF is of 44.0 seconds, with an average of 59.5 seconds. Paradoxically, the receiver will obtain the authenticated fix using satellites that are not transmitting OSNMA.

The discussion about the TTFAF in this section assumes that the navigation data doesn't change, which is true in 97.78\% of the cases (Table \ref{tab:success_iod_optimization}). It also assumes that the cross-authentication tags are transmitted in an optimal sequence from the receiver perspective, which is scenario-specific. Therefore, the values are a lower bound. However, it illustrates how, by enabling the cross-authentication of connected satellites, the performance of the protocol could increase, requiring less satellites in view to obtain an authenticated fix. Transmitting the self-authentication tag \texttt{00S} in the last position of the sequence could also improve the performance by allowing the receivers to start later in the sub-frame and still authenticate the flex tag positions.


\section{CONCLUSION}

Two concrete ideas have been proposed in this paper to improve the TTFAF: page-level processing and COP-IOD optimization. The analysis of the proposed optimizations over three distinct scenarios (Open-Sky, Hard Urban, and Soft Urban) and the test vectors show how the TTFAF can be greatly improved by treating the navigation data received optimally. Moreover, both methods are proven to be complementary when examined in diverse environments. 

The page-level processing for authentication tags and TESLA keys is extremely effective for the urban scenarios, improving the average TTFAF from 127.5 seconds to 94.1 seconds in the Soft Urban scenario and from 266.1 seconds to 151.1 seconds in the Hard Urban scenario. Due to the low satellite visibility and fading, the COP-IOD optimization only improves marginally the average TTFAF for the Soft and Hard Urban scenarios, obtaining 87.5 and 146.1 seconds, respectively.

However, the opposite occurs for the test vectors and the Open-Sky scenario: the page-level processing does not improve the TTFAF, but the COP-IOD optimization reduces it substantially. In both cases, the lack of missed pages inhibits the page-level processing gains. Nonetheless, the COP-IOD optimization benefits from the good satellite visibility of the Open-Sky scenario and the ample number of satellites present in the test vectors. By using this last optimization, the average TTFAF improves from 82.5 seconds to 60.9 seconds for the test vectors and 68.8 seconds for the Open-Sky scenario. The improvement for the lowest TTFAF value is even more impressive, from 68.0 seconds to 44.0 seconds in both cases.

The COP-IOD optimization does not work entirely as expected in the Open-Sky scenario due to the cross-authentication algorithm followed by OSNMA. The algorithm never sends cross-authentication tags for satellites transmitting OSNMA, which generates an imbalance in the number of tags received for each satellite. This behavior adds extra constraints in the minimum number of satellites in view for the COP-IOD optimization to reach lower TTFAF values consistently.

To further improve the OSNMA metrics, it could be useful to implement a multi-frequency library that also uses the I/NAV messages transmitted at E5b. Moreover, Galileo recently implemented four new word types that allow the recovery of missed clock and ephemeris pages using Reed-Solomon encoding, which can significantly improve the performance of OSNMA in urban scenarios \cite{sophie_inav_improvements_2024}.


\section*{ACKNOWLEDGMENT}

The authors would like to thank Sibren De Bast for his insights and comments.


\bibliographystyle{IEEEtran}
\bibliography{biblio.bib}


\end{document}